\documentclass[12pt]{article}
\usepackage{amssymb}
\usepackage{graphics}

\DeclareFontFamily{U}{rsf}{}
\DeclareFontShape{U}{rsf}{m}{n}{
  <5> <6> rsfs5 <7> <8> <9> rsfs7 <10-> rsfs10}{}
\DeclareMathAlphabet\Scr{U}{rsf}{m}{n}

\catcode`\@=11
\def\@citex[#1]#2{%
\if@filesw \immediate \write \@auxout {\string \citation {#2}}\fi
\@tempcntb\m@ne \let\@h@ld\relax \def\@citea{}%
\@cite{%
  \@for \@citeb:=#2\do {%
    \@ifundefined {b@\@citeb}%
      {\@h@ld\@citea\@tempcntb\m@ne{\bf ?}%
      \@warning {Citation `\@citeb ' on page \thepage \space undefined}}%
      {\@tempcnta\@tempcntb \advance\@tempcnta\@ne%
      \@tempcntb\number\csname b@\@citeb \endcsname \relax%
      \ifnum\@tempcnta=\@tempcntb %Number follows previous--hold on to it
        \ifx\@h@ld\relax%
          \edef \@h@ld{\@citea\csname b@\@citeb\endcsname}%
        \else%
          \edef\@h@ld{\ifmmode{-}\else--\fi\csname b@\@citeb\endcsname}%
        \fi%
      \else%   %  non-successor--dump what's held and do this one
        \@h@ld\@citea\csname b@\@citeb \endcsname%
        \let\@h@ld\relax%
      \fi}%
    \def\@citea{,\penalty\@highpenalty\,}%
  }\@h@ld
}{#1}}

\def\@citeb#1#2{{[#1]\if@tempswa , #2\fi}}
\def\@citeu#1#2{{$^{#1}$\if@tempswa , #2\fi }}
\def\@citep#1#2{{#1\if@tempswa , #2\fi}}

\def\bcites{         % cite with []'s
        \catcode`\@=11
        \let\@cite=\@citeb
        \catcode`\@=12
}

\def\upcites{         % cite with exponents
        \catcode`\@=11
        \let\@cite=\@citeu
        \catcode`\@=12
}

\def\plaincites{      % cite without brackets
        \catcode`\@=11
        \let\@cite=\@citep
        \catcode`\@=12
}

\newcount\hour
\newcount\minute
\newtoks\amorpm
\hour=\time\divide\hour by 60
\minute=\time{\multiply\hour by 60 \global\advance\minute by-\hour}
\edef\standardtime{{\ifnum\hour<12 \global\amorpm={am}%
        \else\global\amorpm={pm}\advance\hour by-12 \fi
        \ifnum\hour=0 \hour=12 \fi
        \number\hour:\ifnum\minute<10 0\fi\number\minute\the\amorpm}}
\edef\militarytime{\number\hour:\ifnum\minute<10 0\fi\number\minute}

\def\draftlabel#1{{\@bsphack\if@filesw {\let\thepage\relax
   \xdef\@gtempa{\write\@auxout{\string
      \newlabel{#1}{{\@currentlabel}{\thepage}}}}}\@gtempa
   \if@nobreak \ifvmode\nobreak\fi\fi\fi\@esphack}
        \gdef\@eqnlabel{#1}}
\def\@eqnlabel{}
\def\@vacuum{}
\def\marginnote#1{}
\def\draftmarginnote#1{\marginpar{\raggedright\scriptsize\tt#1}}
\def\draft{
        \pagestyle{plain}
        \overfullrule=2pt
        \oddsidemargin -.5truein
        \def\@oddhead{\sl \phantom{\today\quad\militarytime} \hfil
        \smash{\Large\sl DRAFT} \hfil \today\quad\militarytime}
        \let\@evenhead\@oddhead
        \let\label=\draftlabel
        \let\marginnote=\draftmarginnote
        \def\ps@empty{\let\@mkboth\@gobbletwo
        \def\@oddfoot{\hfil \smash{\Large\sl DRAFT} \hfil}
        \let\@evenfoot\@oddhead}
        \def\@eqnnum{(\theequation)\rlap{\kern\marginparsep\tt\@eqnlabel}%
        \global\let\@eqnlabel\@vacuum}  }

\def\section{\@startsection {section}{1}{\z@}{3.ex plus 1ex minus
 .2ex}{2.ex plus .2ex}{\large\bf}}
\def\subsection{\@startsection{subsection}{2}{\z@}{2.75ex plus 1ex minus
 .2ex}{1.5ex plus .2ex}{\bf}}

\def\appendix{{\newpage\section*{Appendix}}\let\appendix\section%
        {\setcounter{section}{0}
        \gdef\thesection{\Alph{section}}}\section}

\def\abstract{\if@twocolumn
\section*{Abstract}
\else %\small
\begin{center}
{\bf Abstract\vspace{-.5em}\vspace{0pt}}
\end{center}
\quotation
\fi}

\catcode`\@=12

\newcommand{\beq}{\begin{equation}}
\newcommand{\eeq}{\end{equation}}
\newcommand{\beqa}{\begin{eqnarray}}
\newcommand{\eeqa}{\end{eqnarray}}
\newcommand{\dd}{{\rm d}}

\newcommand{\Z}{{\mathbb Z}}

\newcommand{\C}{{\mathbb C}}
\newcommand{\CC}{{\mathbb C}}
\newcommand{\PP}{{\mathbb P}}
\newcommand{\e}{\,{\rm e}}
\newcommand{\CP}{{\CC\PP}}

\newcommand{\be}{\begin{equation}}
\newcommand{\ee}{\end{equation}}
\newcommand{\bea}{\begin{eqnarray}}
\newcommand{\eea}{\end{eqnarray}}

\def\to{\rightarrow}

\def\lae{\mathrel{\mathop{\smash{\lower .5 ex \hbox{$\stackrel<\sim$}}}}}
\def\lae{\mathrel{\mathop{\smash{\lower .5 ex \hbox{$\stackrel>\sim$}}}}}

\def\Tr{{\rm Tr}}

\def\l:{\mathopen{:}\,}
\def\r:{\,\mathclose{:}}

\catcode`\@=11
\def\theequation{\arabic{equation}}
\catcode`\@=12

\bcites

\catcode`\@=11
\def\theequation{\thesection.\arabic{equation}}
\@addtoreset{equation}{section}
\@addtoreset{footnote}{section}
\@addtoreset{footnote}{subsection}
\catcode`\@=12

\newcommand{\opsi}{\overline{\psi}}
\newcommand{\oQ}{\overline{Q}}

\newcommand{\whc}{\widehat{c}}
\newcommand{\nn}{\nonumber}

\newcommand{\ft}[2]{{\textstyle\frac{#1}{#2}}}
\newcommand{\eqn}[1]{(\ref{#1})}
\def\Dslash{\,\,{\raise.15ex\hbox{/}\mkern-12mu D}}
\def\Dbarslash{\,\,{\raise.15ex\hbox{/}\mkern-12mu {\bar D}}}
\def\delslash{\,\,{\raise.15ex\hbox{/}\mkern-9mu \partial}}
\def\delbarslash{\,\,{\raise.15ex\hbox{/}\mkern-9mu {\bar\partial}}}
\def\pslash{\,\,{\raise.15ex\hbox{/}\mkern-9mu p}}
\def\calDslash{\,\,{\raise.15ex\hbox{/}\mkern-12mu {\cal D}}}

\begin{document}
\pagestyle{plain}
\setcounter{page}{1}
\newcounter{bean}
\baselineskip16pt

\begin{titlepage}

\begin{center}
\hfill\today

\vskip 2.3 cm {\Large \bf Aspects of Non-Abelian Gauge Dynamics\\[0.35cm]
in Two-Dimensional ${\mathcal N}=(2,2)$ Theories} \vskip 1.2 cm
{Kentaro Hori$\,{}^1$ and David Tong$\,{}^2$}\\
\vskip 1cm
{\sl ${}^1$ University of Toronto, Ontario, Canada.}\\
{\tt hori@physics.utoronto.ca}
\vskip .3cm
{\sl ${}^2$ DAMTP, University of Cambridge, UK \\
{\tt d.tong@damtp.cam.ac.uk}}

\end{center}

\vskip 0.5 cm
\begin{abstract}
We study various aspects of ${\cal N}=(2,2)$ supersymmetric
non-Abelian gauge theories in two dimensions, with applications to
string vacua. We compute the Witten index of $SU(k)$ SQCD with
$N>0$ flavors with twisted masses; the result is presented as the
solution to a simple combinatoric problem. We further claim that
the infra-red fixed point of $SU(k)$ gauge theory with $N$
massless flavors is non-singular if $(k,N)$ passes a related
combinatoric criterion. These results are applied to the study of
a class of $U(k)$ linear sigma models which, in one phase, reduce
to sigma models on Calabi-Yau manifolds in Grassmannians. We show
that there are multiple singularities in the middle of the
one-dimensional K\"ahler moduli space, in contrast to the Abelian
models. This result precisely matches the complex structure
singularities of the proposed mirrors. In one specific example, we
study the physics in the other phase of the K\"ahler moduli space
and find that it reduces to a sigma model for a second Calabi-Yau
manifold which is not birationally equivalent to the first. This
proves a mathematical conjecture of R{\o}dland.

\end{abstract}

\end{titlepage}

\tableofcontents

\newpage
\section{Introduction}

The purpose of this paper is to study the quantum dynamics of two
dimensional ${\cal N}=(2,2)$ supersymmetric gauge theories with
non-Abelian gauge groups. We will discover a number of interesting
features that are novel to non-Abelian theories and do not occur
for Abelian gauge groups.

Our motivation for this work arose from a number of mathematical
conjectures concerning the moduli spaces of Calabi-Yau manifolds
embedded in Grassmannians \cite{Duco,rodland}. The properties of
these moduli spaces exhibit qualitative differences from those for
Calabi-Yau manifolds in toric varieties. The physicist's tool to
study these conjectures is the linear sigma model, a gauge theory
designed to flow in the infra-red to the desired Calabi-Yau target
space \cite{phases}. While Calabi-Yau manifolds in toric varieties
may be constructed using only Abelian gauge theories, to build
Calabi-Yau in Grassmannians one must necessarily work with
non-Abelian gauge groups.

A typical theory of interest has a $U(k)$ gauge group with $N$
chiral multiplets $\Phi$ in the fundamental representation and a
number of chiral multiplets $P$ which are (negatively) charged
under the central $U(1)\subset U(k)$. The $P$ and $\Phi$ fields
are coupled through a superpotential. The low-energy physics of
this model depends on the value of the Fayet-Iliopolous (FI)
parameter $r$ associated with the central $U(1)\subset U(k)$. For
$r\gg 0$, the fundamental fields $\Phi$ gain an expectation value,
completely breaking the $U(k)$ gauge group and ensuring that one
does not have to contend with strongly coupled non-Abelian
dynamics. It is in this regime that the gauge theory reduces to a
sigma-model on the compact Calabi-Yau 3-fold of interest. However,
to fully understand the moduli space of the Calabi-Yau one must
also study the theory at small values of $r$ and at $r\ll 0$. In
these regions, the full force of the non-Abelian gauge dynamics is
at play. For example, at $r=0$ there is a locus on which the
$U(k)$ gauge group is unbroken and a $k$ dimensional non-compact
Coulomb branch with an unbroken $U(1)^k$ emerges. Similarly, for
$r\ll 0$, the $P$ fields gain an expectation value, breaking
$U(k)$ to $SU(k)$.

We are therefore invited to study the dynamics of $U(k)$ and
$SU(k)$ gauge theories coupled to fundamental chiral multiplets.
We answer basic questions concerning the vacuum structure of these
theories: How many supersymmetric ground states are there? Under
what circumstances are the ground states normalizable? When does
the theory flow to a non-trivial fixed point?
In answering these questions, we find
several new phenomena which do not occur for Abelian models. We
now give a summary of the main results:

\medskip
\noindent
{\bf The Witten Index:}

Although the Witten index \cite{index}
for $U(k)$ theories in two dimensions is easy to compute, to our
knowledge the calculation for $SU(k)$ theories has not appeared in
the literature. In Section~\ref{sec:index} we derive the Witten
index for ${\cal N}=(2,2)$ $SU(k)$ supersymmetric QCD with $N$
flavors, each endowed with a twisted mass \cite{HH}. As we will
explain, the non-compact Coulomb branch of this theory is lifted
by quantum effects, leaving behind isolated, supersymmetric vacua.
The Witten index is given by the solution to a simple combinatoric
problem: Find $k$ distinct $N$-th roots of unity, modulo overall
scaling, whose sum is non-zero. In particular, there is no
supersymmetric ground state for $1\leq N\leq k$ and there is
exactly one for $N=k+1$. We list below the index for low values of
$k$ and $N$.
\begin{center}
\begin{tabular}{|c||c|c|c|c|c|c|c|c|c|c|c|c|c|c|c|}
\hline
$k\setminus N$
  &1&2&3& 4& 5& 6& 7& 8& 9&10&11&12& 13& 14& 15\\
\hline\hline
2 &0&0&1& 1& 2& 2& 3& 3& 4& 4& 5& 5&  6&  6&  7\\
\hline
3 &0&0&0& 1& 2& 3& 5& 7& 9&12&15&18& 22& 26& 30\\
\hline
4 &0&0&0& 0& 1& 2& 5& 8&14&20&30&40& 55& 70& 91\\
\hline
5 &0&0&0& 0& 0& 1& 3& 7&14&25&42&65& 99&143&200\\
\hline
6 &0&0&0& 0& 0& 0& 1& 3& 9&20&42&75&132&212&333\\
\hline
7 &0&0&0& 0& 0& 0& 0& 1& 4&12&30&65&132&245&429\\
\hline
\end{tabular}
\\
[0.2cm]{\bf Table 1:} The Witten index for $SU(k)$ SQCD with $N$
massive flavors
\end{center}

\medskip
\noindent
{\bf IR Dynamics of $SU(k)$ Gauge Theories:}

 In
Section~\ref{sec:IR}, we study the infra-red dynamics of $SU(k)$
gauge theories with $N$ {\it massless} fundamental chiral
multiplets. Such theories are expected to
flow to superconformal field theories (SCFTs) in the infra-red limit.
For example, if the superpotential is a homogeneous polynomial of degree $d$
in the baryon operators,
the gauge theory flows to a SCFT
with central charge $\whc=c/3=N(k-2/d)-k^2+1$.
However, the potential existence of a Coulomb branch --- a
non-compact, flat direction in field space
--- means that the ground state wavefunction may spread, rendering
the conformal field theory singular. Such behavior is seen at the
conifold point of ${\cal N}=(2,2)$ Abelian theories \cite{phases}
and at the special point of ${\mathcal N}=(4,4)$ SQED
\cite{comments,Higgs,AB}. We propose that the $SU(k)$ theory
suffers from such a singularity if and only if there exist $k$
distinct $N$-th roots of unity that sum to zero.
For example,
the low energy theory of
an $SU(k)$ theory with
$N=k+1$ fundamentals is always non-singular and
is described by the $N$ baryon operators as the independent variables.
In particular, $SU(k)$ SQCD with $N=k+1$ massless flavors
flows to a free conformal field theory with $\whc=N$.
We also propose an infra-red duality between the $SU(k)$ gauge theory and
the $SU(N-k)$ gauge theory, both with $N$ fundamentals and
degree $d$ superpotential for the baryons.
The duality is proved for the case ${2N\over Nk-k^2+1}<d\leq N$.

\medskip
\noindent
{\bf Splitting the Conifold Singularity:}

 For the class of $U(k)$
linear sigma models described above, we find that there are
typically multiple singular points in the middle of the
one-dimensional K\"ahler moduli space parameterized by the
FI-theta parameter $t=r-i\theta$. This behavior is in contrast to
$U(1)$ theories which always have exactly one singular point in a
one-dimensional K\"ahler moduli space. These multiple
singularities, which  arise from a quantum mechanical splitting of
the classical singularity, are the topic of
Section~\ref{sec:split}. The analysis is based on the the quantum
potential on the Coulomb branch \cite{phases,verlinde}. If there
are $N$ fundamental chiral multiplets, and several fields charged
under the $U(1)\subset U(k)$, the number of singular points
coincides with the Witten index for $SU(k)$ SQCD with $N$ massive
flavors. At each of these points, a one-dimensional subspace in
the $k$-dimensional Coulomb branch becomes a truly flat direction.
For the models corresponding to Calabi-Yau three-folds, the result
precisely matches the singularities of the complex structure
moduli space of the proposed mirrors, giving a strong support to
the mirror symmetry conjecture of \cite{Duco}.

\medskip
\noindent
{\bf The Glop Transition:}

 One of the original motivations for
our work is the conjecture by R{\o}dland \cite{rodland} that two
inequivalent compact Calabi-Yau three-folds $X$ and $Y$ sit on the
same one-dimensional complexified K\"ahler moduli space. $X$ is a
submanifold in the Grassmannian $G(2,7)$ while $Y$ is an {\it
incomplete} intersection in $\CP^6$, referred to as the Pfaffian
Calabi-Yau. We wish to understand this claim from a purely quantum
field theoretic point of view. The natural linear sigma-model for
the Calabi-Yau $X$ is of the type described above: it is a $U(2)$
gauge theory with $7$ fundamental chiral multiplets $\Phi_i$ and a
further $7$ chiral multiplets $P_i$, transforming in the
$\det^{-1}$ representation (i.e. charged only under the central
$U(1)\subset U(2)$). These chiral multiplets are coupled through
the superpotential
\be W =
\sum_{i,j,k=1}^7A_i^{jk}\,P^i(\Phi^1_j\Phi^2_k-\Phi^2_j\Phi^1_k)
\label{starting}
\ee
where $A^{jk}_i$ are generic coefficients that are anti-symmetric
in the upper indices. In the regime $r \gg 0$, the manifold $X$
appears, sitting at the bottom of the scalar potential of this
theory. Moving towards the center of the K\"ahler moduli space,
around $r\sim 0$, one finds three singular points --- this can be
read from the $(k,N)=(2,7)$ entry of Table 1 for the Witten index.

The regime $r\ll 0$ is the focus of Section~\ref{sec:Rodland},
where we make use of many results from previous sections. The
D-term equations force the $P$'s to span $\CP^6$ and the low
energy theory includes an unbroken $SU(2)$ gauge theory with $N=7$
flavors $\Phi_i$. The superpotential \eqn{starting} provides a
complex mass matrix $A(P)^{jk}=\sum_{i=1}^7A_i^{jk}P^i$ for these
flavors which varies as we move around $\CP^6$. The
antisymmetric matrix $A(P)$ has rank $6$ at a generic point $P$ of
$\CP^6$ but it degenerates to rank $4$ at a locus of codimension
three. This rank $4$ locus is precisely the location of the
Pfaffian Calabi-Yau $Y$. The number of massless flavors is $N=1$
on the rank $6$ domain but it jumps to $N=3$ on the rank $4$
locus. Looking at the Witten index in Table 1, we see that
supersymmetry is broken on the rank $6$ domain and the low energy
theory localizes on the rank $4$ locus. In addition to the tangent
modes, the low energy theory on the rank $4$ locus contains an
$SU(2)$ gauge multiplet, $N=3$ massless fundamentals, and three
singlets from the transverse modes of $P$. However, since any
$SU(k)$ gauge theory with $N=k+1$ fundamentals is described at low
energies by the $N$ independent baryon variables, this extra
sector with interaction (\ref{starting}) flows to a Landau-Ginzburg model
with a non-degenerate and quadratic superpotential 
for the three baryons and the three transverse modes. 
It has a unique
supersymmetric ground state with a mass gap, and hence only the
tangent modes to the rank $4$ locus remain in the deep infra-red
limit. In this way we obtain the supersymmetric non-linear sigma
model on the Pfaffian Calabi-Yau $Y$. This provides a field
theoretic proof of R{\o}dland's conjecture \cite{rodland}. Note
that the Calabi-Yau $Y$ does not arise from simply restricting to
the zeroes of a classical potential, but requires us to make use
of the strongly coupled non-Abelian gauge dynamics in a novel
fashion.

The transition from the $X$ phase to the $Y$ phase is smooth
within the conformal field theory. This is similar to the
transition from large volume Calabi-Yau to a Landau-Ginzburg
orbifold or, perhaps more closely, to the familiar flop transition
\cite{phases,flop}. However, our example exhibits one crucial
property that distinguishes it from the flop: $X$ and $Y$ are {\it
not} birationally equivalent. For this reason we refer to this new
type of topology changing transition as a Grassmannian flop, or
glop.
\begin{figure}[htb]
\centerline{\includegraphics{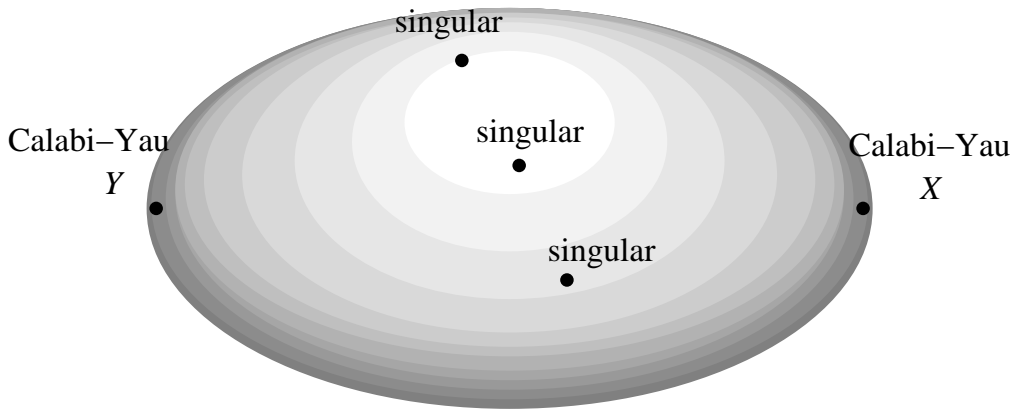}} \caption{The K\"ahler
Moduli Space of the IR fixed points of the theory with superpotential
(\ref{starting}). It has two large volume limits and
three singularities} \label{Rodland}
\end{figure}

\section{Splitting the Conifold Singularity}
\label{sec:split}

In this section we discuss the phenomenon of the quantum splitting
of conifold singularities for Calabi-Yau manifolds
embedded in Grassmannians. The basic physics can already be
seen in non-compact Calabi-Yau, constructed from line-bundles over
Grassmannians. We start with a description of the relevant gauged
linear sigma models, before presenting a derivation of our result.
We then compare these results with predictions from mirror
symmetry for compact Calabi-Yau 3-folds and find agreement in all
cases.

\subsection{The Models}

Throughout this paper, our main tool will be the gauged linear
sigma model, a gauge theory designed to flow in the infra-red to a
non-linear sigma model on a target space ${\cal M}$ which arises
as the classical vacuum moduli space of the theory \cite{phases}.

We will work with a $U(k)$ gauge group whose field strength
$F_{01}$ lives in a twisted chiral multiplet $\Sigma$. Our
notation is canonical and follows \cite{phases,book},
\beq
\Sigma=\sigma+\theta^+\bar{\lambda}_++\bar{\theta}^-\lambda_++
\theta^+\bar{\theta}^-(D-iF_{01})+\ldots\ \ . \eeq
The gauge field couples to $N$ chiral multiplets $\Phi_i$
transforming in the fundamental representation ${\bf k}$. Each has
a component expansion,
\beq
\Phi_i=\phi_i+\theta^+\psi_{i+}+\theta^-\psi_{i-}+\theta^+\theta^-F_i+\ldots
\ \ \ \ \ \ \ i=1,\ldots,N\ \ .\eeq
To ensure that the theory flows to an interacting conformal field
theory --- or, equivalently, to ensure that ${\cal M}$ is
Calabi-Yau --- we need to add further matter to cancel the axial
$U(1)_A$ anomaly \cite{phases}. This can be achieved in at least
two different ways: we could simply introduce $N$ companion chiral
fields $\tilde{\Phi}_i$, each transforming in the anti-fundamental
representation $\bar{\bf k}$ of the gauge group. However, a more
interesting possibility is to instead add $S$ chiral multiplets
$P^\alpha$, each transforming in the $\det^{-q_\alpha}$
representation for some choice of integers $q_\alpha$. This means
that the field $P^\alpha$ is charged only under the central
$U(1)\subset U(k)$ so that for $g\in U(k)$, $P^\alpha\rightarrow
(\det^{-q_\alpha}g)P^\alpha$ or, infinitesimally, for
$g=1+\epsilon$,  $\delta P^\alpha =-q_\alpha(\Tr
\epsilon)P^\alpha$. We write
\beq
P^\alpha=p^\alpha+\theta^+\chi^\alpha_{+}+\theta^-\chi^\alpha_{-}
+\theta^+\theta^-F^\alpha+\ldots\ \ . \eeq
The condition for $U(1)_A$ axial anomaly cancellation is
\beq N\Tr_{\bf k}\, F_A +
\sum_{\alpha=1}^S\Tr_{\det^{-q_\alpha}}\,F_A
=\left(N-\sum_{\alpha=1}^Sq_\alpha\right)\Tr_{\bf k}\,F_A = 0\ ,
\eeq
which requires $\sum_{\alpha=1}^Sq_{\alpha}=N$.

The $U(k)$ adjoint valued $D$-term for the theory reads,
\beq D^a_{\ b}= e^2 \left(\sum_{i=1}^N \phi^a_i\phi_{bi}^\dagger -
\sum_{\alpha=1}^Sq_\alpha|p^\alpha|^2\delta^a_{\ b}-r\delta^a_{\
b}\right)\ \ \ \ a,b=1,\ldots, k\ . \label{D}\eeq
where $r$ is the Fayet-Iliopoulos (FI) parameter and $e^2$ the
gauge coupling constant. In the absence of a superpotential, the
vacuum moduli space is ${\cal M}\cong\{D=0\}/U(k)$, a non-compact
Calabi-Yau manifold which, for $r>0$, is the sum of $S$ line
bundles over the Grassmannian $G(k,N)$ of $k$-planes in $\C^N$: 
${\cal M}\cong\oplus_\alpha\, {\cal O}(q^\alpha)\rightarrow
G(k,N)$.

We can construct compact Calabi-Yau manifolds $X\subset {\cal M}$
through the introduction of a superpotential for the chiral
multiplets \cite{phases}. We consider gauge invariant
superpotentials of the form,
\beq W=\sum_{\alpha=1}^S P^{\alpha}G_{\alpha}(B), \eeq
where $G_{\alpha}$ is a polynomial of degree $q_{\alpha}$ in the
Pl\"ucker coordinates (i.e. baryonic variables)
\beq B_{i_1...i_k}
=\epsilon_{a_1...a_k}\Phi^{a_1}_{i_1}\cdots \Phi^{a_k}_{i_k} \eeq
These provide homogeneous coordinates on the projective space
$\PP(\wedge^k\C^N)\cong \CP^{{N\choose k}-1}$ in which the
Grassmannian is embedded. For $r\gg0$, provided certain genericity
conditions on $G_\alpha$ hold, the low-energy theory is the
non-linear sigma model on the compact Calabi-Yau
$X_{q_1,\ldots,q_S}\subset G(k,N)$ of dimension $Nk-k^2-S$,
defined by the intersection of hypersurfaces $G_\alpha=0$ with the
Grassmannian $G(k,N)$ living at $p^\alpha=0$. We will discuss the
precise genericity conditions on $G_\alpha$ for a specific example
in Section~\ref{sec:Rodland}.

If we restrict attention to 3-folds, the above construction yields
only a handful of compact Calabi-Yau manifolds. Indeed, the
dimensionality condition $Nk-k^2-S=3$ can be written as
$$
(k-1)(N-k-1)=4+S-N,
$$
but the right hand side is at most $4$
since we have $S\leq N$ from the Calabi-Yau condition
$N=\sum_{\alpha=1}^Sq_{\alpha}$.
We list the solutions below, together
with their relevant Hodge numbers (taken from \cite{Duco}).
\begin{center}
\begin{tabular}{|c|c|c|} \hline
$X$ & $h^{1,1}(X)$ & $h^{2,1}(X)$ \\ \hline $X_4\subset G(2,4)$ &
1 & 89  \\ \hline $X_{1,1,3}\subset G(2,5)$ & 1 & 76  \\ \hline
$X_{1,2,2}\subset G(2,5)$ & 1 & 61  \\ \hline
$X_{1,1,1,1,2}\subset G(2,6)$ & 1 & 59  \\ \hline
$X_{1,\ldots,1}\subset G(2,7)$ & 1 & 50  \\ \hline
$X_{1,\ldots,1}\subset G(3,6)$ & 1 & 49  \\ \hline
\end{tabular}\\
[0.2cm] {{\bf Table 2:}~ 3-fold complete intersections in
Grassmannians.}
\end{center}
The sole K\"ahler modulus of each manifold $X$ is inherited from
the Grassmannian in which $X$ is embedded. The moduli space of the
complexified K\"ahler class is parameterized by the complex
combination of the FI parameter $r$ and the theta angle $\theta$.
\beq t=r-i\theta. \eeq
For generic $t$, the conformal field theory on $X$ has
well-defined correlation functions. However at certain values of
$t$, the correlation functions diverge. From the perspective of
the gauge theory, this divergence can be traced to the emergence
of a new massless direction in field space --- the Coulomb branch
--- into which the ground state wavefunctions spread
\cite{phases}. Let us firstly recall how one sees the emergence of
the Coulomb branch in Abelian gauge theories.

\subsection{Abelian Theories}
\label{subsec:abe}

Consider a $U(1)$ gauge theory with $N$ chiral multiplets of
charge $Q_i$, $i=1,\ldots N$. The criterion for conformal
invariance is $\sum_iQ_i=0$. Since the location of the
singularities in the K\"ahler moduli space does not depend on the
complex structure moduli, we may perform the analysis for the
situation with vanishing superpotential and the corresponding
non-compact Calabi-Yau. In this case the classical vacuum energy
is given by,
\beq V=\frac{e^2}{2}\left(\sum_{i=1}^N
Q_i\left|\phi_i\right|^2-r\right)^2 + \sum_{i=1}^N
Q_i^2|\sigma|^2\left|\phi_i\right|^2 \ \ .\eeq
For all non-zero finite values of $r$, the classical vacuum
equation $V=0$ results in a Higgs branch of vacua corresponding to
a non-compact Calabi-Yau manifold while the vector multiplet
scalar is forced to vanish: $\sigma=0$. However, for $r=0$ the
classical Coulomb branch opens up, parameterized by non-zero
$\sigma$ while the chiral multiplet scalars are now forced to
vanish: $\phi_i=0$.

To see whether the Coulomb branch survives in the quantum theory,
we  compute the exact potential energy in the large $|\sigma|$
region of the field space. All charged matter is heavy and may be
safely integrated out, resulting in an effective twisted
superpotential $\tilde{W}$
\cite{dadda,phases,mp}\footnote{Throughout the paper, we use the
convention of \cite{hv,book} in which the action is multiplied by
$1\over 2\pi$ in the exponent of the path-integral weight.}:
\bea
\tilde{W}&=&
-t\Sigma-\sum_{i=1}^NQ_i\Sigma\left(\,\log(Q_i\Sigma)-1\right)
\nn\\
&=& -\Sigma\left(t+\sum_{i=1}^NQ_i\log Q_i\right), \eea
where $\sum_iQ_i=0$ is used in the last step. The lowest energy
density is given by \cite{phases,coleman} \beq U(\sigma)={e_{\it
eff}(\sigma)^2\over 2} \min_{n\in \Z}\left|\ t+\sum_{i=1}^NQ_i\log
Q_i+2\pi in\ \right|^2, \eeq where $e_{\it eff}(\sigma)$ is an
effective gauge coupling that approaches the bare value $e$ as
$|\sigma|\to +\infty$. Thus the truly flat Coulomb branch appears
when
\beq t=-\sum_{i=1}^NQ_i\log Q_i \ \ \ \mbox{(modulo $2\pi
i\Z$)}.\label{newman}\eeq
The main quantum effect is the contribution of the $2\pi$ periodic
theta angle $\theta$; a secondary effect is the finite shift of
the parameters. The end result is that there is a {\it single}
value of the complex FI parameter $t$ for which the Coulomb branch
emerges. At this point, correlation functions of Higgs branch
operators diverge. For this reason the value of $t$ \eqn{newman}
is referred to as a {\it singular} point in the one-dimensional
quantum K\"ahler moduli space. In the dual mirror theory, it
corresponds to the point of the complex structure moduli space
where the mirror manifold develops a conifold singularity
\cite{quintic,hv,hiv}. By abuse of language, we will also refer to
the singular point in the K\"ahler moduli space as the ``conifold
point''.

In models with more than one K\"ahler moduli, the multi-dimensional
 FI-parameter space is separated into various ``phases''.
With the inclusion of the theta angles, the ``phase boundaries''
lift up to a subvariety of complex codimension one where the
theory is singular. (See, for example, figure 1-3 of \cite{mp} for
a graphical representation of this in a specific example). Each
phase boundary has an asymptotic region where the unbroken gauge
group is one $U(1)$ and there too the number of singular loci is
{\it exactly one}.

\subsection{Non-Abelian Theories}
\label{subsec:na}

We would now like to repeat this analysis for non-Abelian gauge
theories. As we shall see, quantum corrections to the classical
singularity are much more pronounced. We consider a $U(k)$ gauge
theory with $N$ chiral multiplets $\Phi_i$ transforming in the
fundamental representation ${\bf k}$, and a further $S$ chiral
multiplets $P_\alpha$ transforming in the $\det^{-q_\alpha}$
representation. Again we turn off the superpotential, $W=0$, as it
does not affect the singularity analysis. As discussed in
Section~\ref{subsec:abe}, the criterion for conformal invariance
is $\sum_\alpha q_\alpha=N$. The classical potential is
\bea V&=&{1\over 2e^2}\Tr\, [\sigma,\sigma^\dagger]^2
+\frac{e^2}{2}\Tr\,\left(\sum_{i=1}^N \phi_i\phi_{i}^\dagger -
\sum_{\alpha=1}^Sq_{\alpha}|p^{\alpha}|^2{\bf 1}_k -r{\bf
1}_k\right)^2 \nn\\ \ \ \ && +{1\over
2}\sum_{i=1}^N\phi^\dagger_i\{\sigma^\dagger,\sigma\}\phi_i +
\sum_{\alpha=1}^Nq_{\alpha}^2\left|\Tr\,\sigma\right|^2|p^{\alpha}|^2.
\eea
The classical vacuum equation $V=0$ first of all requires that
$[\sigma,\sigma^\dagger]=0$, that is, $\sigma$ is diagonalizable,
\beq \sigma=\left(\begin{array}{ccc}
\sigma_1&&\\
&\ddots&\\
&&\sigma_k
\end{array}\right).
\label{diago} \eeq
The eigenvalues are generically all distinct and the $U(k)$ gauge
group is broken to its maximal torus $U(1)^k$. However, as some of
the eigenvalues coalesce there is an enhanced unbroken non-Abelian
subgroup. The classical theory has different Coulomb branches for
different values of $r$. For $r>0$, the D-term equation requires
that $\phi\phi^{\dag}$ has maximal rank $k$, breaking the entire
$U(k)$ gauge group: the Coulomb branch is completely lifted
($\sigma =0$) and we are left only with the Higgs branch which is
a smooth non-compact Calabi-Yau space ${\cal M}$ of dimension
$Nk+S-k^2$. In contrast, for $r<0$, the $p$'s must have non-zero
values which break only the central $U(1)\subset U(k)$ subgroup.
The other scalars $\phi$ can have various ranks, from $0$ to $k$,
 and the residual $SU(k)$ gauge symmetry
is broken to the complementary subgroups. The classical vacuum
manifold at $r<0$ is thus a mixed Coulomb-Higgs branch: Coulomb
branches of various dimensions, from $0$ to $(k-1)$, sit over loci
of the Higgs branch classified by the rank of $\phi$. Finally, for
$r=0$ the full $k$-dimensional Coulomb branch emerges at
$\phi=p=0$.

Let us examine the existence of a genuine quantum Coulomb branch
by computing the exact effective potential. We assume that
$\sigma$ is diagonalizable (\ref{diago}), and furthermore
\beqa
&&\sigma_a\ne\sigma_b\ \ \mbox{if $a\ne b$,}\nn\\
[0.2cm]&&\sigma_a\ne 0,\ \ \forall\ a,
\label{validity}\\
&&\sum_{a=1}^k\sigma_a\ne 0\nn \eeqa
to suppress all non-trivial interactions. Then all the charged
multiplets and W-bosons are heavy and can be integrated out. This
results in the effective twisted superpotential,
\beq \tilde{W}=-t\sum_{a=1}^k\Sigma_a
-\sum_{a=1}^kN\Sigma_a\left(\log\Sigma_a-1\right)
+\sum_{\alpha=1}^Sq_\alpha\left(\sum_{a=1}^k\Sigma_a\right)\left[
\log\left( -q_\alpha\sum_{a=1}^k\Sigma_a\right)-1\right].
\label{tildew}\eeq
Note that there is no contribution from the W-boson integrals
\cite{verlinde}. Let us see whether there is a true flat direction
within the validity range (\ref{validity}) of this superpotential.
First we show that with a fixed trace
\beq \sum_{a=1}^k\Sigma_a=\mbox{fixed}\, \eeq
the remaining $(k-1)$ relative modes are massive. To see this we
add the Lagrange multiplier term
\beq \Delta
\tilde{W}=\lambda\left(\Sigma-\sum_{a=1}^k\Sigma_a\right)
\label{lm}\eeq
and extremize $\tilde{W}+\Delta \tilde{W}$ with respect to
$\lambda$ as well as the independent $\Sigma_a$'s:
\beq \frac{\partial(\tilde{W}+\Delta\tilde{W})}{\partial \Sigma_a}
= -N\log\Sigma_a +\sum_{\alpha=1}^Sq_\alpha\log(-q_\alpha \Sigma)
-(t+\lambda)=0. \label{exsup}\eeq
For a fixed value of $\Sigma$, the
components $\Sigma_a$ are constrained to obey
$\Sigma_a^N=\exp(\Xi)$ where the quantity $\Xi=\sum_\alpha
q_\alpha\log(-q_\alpha\Sigma)-(t+\lambda)$ is independent of the
gauge index $a$. The solutions to equation \eqn{exsup} take the
form,
\beq
\Sigma_a=\frac{\omega^{n_a}}{Z}\,\Sigma
\label{sigsol}
\eeq
where $\omega=\e^{2\pi i /N}$ and $n_a\in\{0,1,\ldots,N-1\}$ are
some choice of $k$ integers and
\beq Z=\sum_{a=1}^k\omega^{n_a}.\eeq
These solutions do not come in continuous families, but are
isolated: this is the statement that the relative $\Sigma_a$'s are
massive in the background of a fixed trace $\Sigma$. For the
solution (\ref{sigsol}) to lie in the range of validity
(\ref{validity}), we require that the $k$ choices ${n_a}$ are all
distinct and that $Z=\sum_a\omega^{n_a}$ is non-vanishing. Note
also that the uniform shift $n_a\to n_a+1$ $\forall\ a$ results in
$Z\to \omega Z$ and does not change the ratio $\Sigma_a$. Finally,
a permutation of the $n_a$'s is a gauge transformation.

To summarize: the number of trustworthy solutions for $\Sigma_a$
given by equation \eqn{sigsol} is equal to the number of distinct
choices of $k$ integers $n_a$ from $\{0,1,\ldots,N-1\}$, modulo
shifts $n_a\rightarrow n_a+1$, with the requirement that
$Z=\sum\omega^{n_a}\neq 0$. Let us denote this number as $n(k,N)$.
It will play an important role in the following section.

Finally, we turn to the original question: when does a Coulomb
branch for $\Sigma$ arise? We may substitute the solution
\eqn{sigsol} into the effective superpotential \eqn{tildew} to
find,
\beq \tilde{W}=\Sigma\left(-t+N\log
Z+\sum_{\alpha=1}^Sq_\alpha\log(-q_\alpha)\right) \eeq
So, as with the Abelian case, we find that the Coulomb branch
exists when the FI-theta parameter takes the specific values,
\beq \prod_\alpha q_\alpha^{\,q_\alpha}\e^{-t} =\frac{(-1)^N}{Z^N}
\label{sigsing}\eeq
Only a one-dimensional Coulomb branch emerges from each of these
points, rather than a $k$-dimensional branch that one might expect
from a $U(k)$ gauge theory. Each Coulomb branch is parameterized
by the scalar $\Sigma =\sum_a\Sigma_a$ associated to the quotient
group $U(1)=U(k)/SU(k)$. The different branches are characterized
by the mass spectrum of the W-bosons, given by
$\Sigma_a-\Sigma_b$, where $\Sigma_a$ satisfy \eqn{sigsol}.

For our immediate purposes, the crucial point is that the value of
$t$ for which the one-dimensional Coulomb branch exists is not
unique; it exists for every non-vanishing value of
$Z=\sum\omega^{n_a}$ with distinct integers
$n_a\in\{0,1,\ldots,N-1\}$. Thus, in general, there are multiple
singular points in the middle of the K\"ahler moduli space.

We have not yet analyzed the quantum fate of the mixed
Coulomb-Higgs branches that exist classically at any negative
value of $r$. We will show in Section~\ref{subsec:inter} that
these are lifted and do not introduce further singularities in the
K\"ahler moduli space.

\subsection{Comparison with Mirror Symmetry Conjecture}
\label{subsec:mirror}

In \cite{Duco} Batyrev, Ciocan-Fontanine, Kim and van Straten
constructed the mirror manifold $Y$ for each Calabi-Yau 3-fold $X$
listed in Table 1. The solution to the Picard-Fuchs equation
provides the classical Yukawa coupling $K^{(3)}_z$ of the mirror
theory in terms of the single complex structure $z$. In this
section we check that the discriminant locus of $Y$ coincides with
the quantum singularities \eqn{sigsing} found above. We
deal with each example in turn, writing down the Yukawa coupling
computed in \cite{Duco} before comparing to the quantum
singularities:

\begin{itemize}
\item The simplest 3-fold $X_4\subset G(2,4)$ may also be written
as $V_{2,4} \subset \CP^5$, the intersection of a quadric, which
restricts to $G(2,4)\subset \CP^5$, with a quartic. The mirror was
previously computed in \cite{batstrat} and \cite{vt} (see example
6.4.2 of the former paper).

Since this example may be rewritten as an Abelian model, we do not
expect the singularity to split. Indeed, the correlation
functions of the mirror have a single pole at $z=2^{-10}$
\cite{Duco}. This is to be compared with equation \eqn{sigsing}
which yields singularities when $4^4\e^{-t}=1/Z^4$. For this
example, there is a unique choice: $Z=1+\omega$. (Of the other
choices, $Z=1+\omega^2=0$ so is invalid, while $Z=1+\omega^3$ is
related to the first choice by a shift $Z\rightarrow \omega^3 Z$).
The location of the pole agrees with the mirror under the
identification $z=-\e^{-t}$.

\item For the next example $X_{1,1,3}\subset G(2,5)$, the
computation of Batyrev et al. yields
\beq K_z^{(3)} = \frac{15}{1-11(3^3z)-(3^3z)^2} \label{x113}\eeq
The Coulomb branch analysis \eqn{sigsing} gives singularities when
$3^3\e^{-t}=-1/Z^5$. This time there are two possible choices with
$Z\neq 0$. They are $Z=1+\omega$ and $Z=1+\omega^2$, yielding
singularities at $3^3\e^{-t}=-11/2\pm5\sqrt{5}/2$ in agreement with
the mirror symmetry prediction \eqn{x113}. Note that this tells us
we should identify the coordinates as $z=\e^{-t}$.

\item The mirror of the Calabi-Yau $X_{1,2,2}\subset G(2,5)$ has
Yukawa coupling
\be K_z^{(3)} = \frac{20}{1-11(2^4z)-(2^4z)^2} \ee
Comparing the denominator to \eqn{x113}, we see that the position
of the singularities remains the same up to the normalization of
$z$. Indeed, we see the same behavior in our quantum theory where
the normalization arises from
$\prod_{\alpha}{q}_\alpha^{q_{\alpha}}=2^4$. Once again find
agreement if we identify $z=\e^{-t}$.

\item The mirror of the Calabi-Yau $X_{1,1,1,1,2}\subset G(2,6)$
has Yukawa coupling
\be K_z^{(3)} = \frac{28}{((2^2z)+1)(27(2^2z)-1)} \ee
From \eqn{sigsing}, the singularities occur at
$(2^2)\e^{-t}=1/Z^6$. There are two possibilities: $Z=1+\omega$ and
$Z=1+\omega^2$. This gives $Z^6=-27$ and $1$ respectively,
resulting in agreement for $z=-\e^{-t}$

\item The mirror of the Calabi-Yau $X_{1,\ldots,1}\subset G(2,7)$
has Yukawa coupling
\be K_z^{(3)} = \frac{42-14z}{1-57z-289z^2+z^3} \ee
The quantum singularities of \eqn{sigsing} occur at
$\e^{-t}=-1/Z^7$, where there are now three choices: $Z=1+\omega$,
$Z=1+\omega^2$ and $Z=1+\omega^3$. It is a simple matter to check
numerically that these reproduce the three poles of the Yukawa
coupling with $z=\e^{-t}$.

\item Finally, the mirror of the Calabi-Yau $X_{1,\ldots,1}\subset
G(3,6)$ has Yukawa coupling\footnote{The original formula of
\cite{Duco} contains a minus sign typographical error in the denominator.
We thank Duco van Straten for confirmation of this.}
\be K_z^{(3)} = \frac{42}{(1-z)(1-64z)} \label{xg36}\ee
The singularities from \eqn{sigsing} lie at $\e^{-t}=1/Z^6$. This
time we have {\it three} choices for $Z$. They are
$Z=1+\omega+\omega^2$, $Z=1+\omega+\omega^3$ and
$Z=1+\omega+\omega^4$. They yield $Z^6=64,1,1$ respectively. Note
that the presence of the two solutions with $Z=1$ appears not to
lead to a double pole in the Yukawa coupling. Once again, the
dictionary between $t$ and the complex structure $z$ chosen in
\cite{Duco} requires $\e^{-t}=z$ for agreement.

\end{itemize}
It is worth noting that for each example, the map takes the form
$\e^{-t}=(-1)^{k+N+1}\,z$. Note that $z$ is chosen in \cite{Duco}
so that it approaches $\e^{-T}$ in the large volume limit, where
$T$ is the complexified K\"ahler class $T=\int_{C}(\omega-iB)$.
The sign $(-1)^{N+k+1}$ shows that the linear sigma model theta
angle is asymptotically related to the
B-field by
\beq \int_CB\ \simeq\  \theta+\pi(N+k+1)\ \
\mbox{modulo $2\pi \Z$}.
\eeq
It would be interesting to
understand the origin of this shift.
(See \cite{mp} for a discussion of this point in Abelian models.)

\section{The Witten Index for Two Dimensional SQCD}
\label{sec:index}

In this section we calculate the Witten index for ${\cal N}=(2,2)$
$SU(k)$ gauge theory with $N$ chiral multiplets in the fundamental
 representation ${\bf k}$. To our knowledge,
a computation of the index has not previously appeared in the
literature. Recall that for a $U(k)$ gauge group coupled to $N$
fundamental chiral multiplets there are ${N\choose k}$ vacua. The
question is: what happens when we decouple the overall $U(1)$?

Theories with extended supersymmetry typically have non-compact
moduli spaces of vacua making the computation of the Witten index
tricky at best, ill-defined at worst \cite{index}. Our model is no
exception. When the chiral multiplets have vanishing mass, the
classical theory has both a non-compact Higgs branch of complex
dimension $k(N-k)+1$ and a non-compact Coulomb branch of dimension
$k-1$. In this section, we deform the theory  by endowing the
chiral multiplets with twisted masses \cite{HH}. As we shall see,
generic masses lift both Higgs and Coulomb branches and render the
Witten index well-defined. We now compute this index, which we
call $w(k,N)$. 
We postpone the discussion of massless flavors to Section~\ref{sec:IR}.

\subsection{A First Attempt at Counting}
\label{subsec:count}

We start with a direct approach. This will give the right
answer despite a number of shortcomings which we later
remedy by taking a more oblique route.
 Let us give the twisted mass $\tilde{m}_i$ to the $i$-th
fundamental field $\Phi_i$. The scalar potential for the $SU(k)$
theory is
\bea V&=&{1\over 2e^2}\Tr|[\sigma,\sigma^\dagger]|^2
+\frac{e^2}{2}\Tr\,\left[\sum_{i=1}^N\left(
\phi_i\phi_{i}^{\dag}-{1\over k} \Tr(\phi_i\phi_{i}^{\dag}){\bf
1}_k\right)\right]^2 \nn\\ \ \ \ && +{1\over
2}\sum_{i=1}^N\phi^\dagger_i\{\sigma^\dagger
-\tilde{m}_i^*,\sigma-\tilde{m}_i\}\phi_i. \eea The vacuum
equation $V=0$ requires $[\sigma,\sigma^{\dag}]=0$, the D-term
equation \beq \phi\phi^{\dag}={1\over k}\Tr(\phi\phi^{\dag}){\bf
1}_k, \label{Deq} \eeq for the $k\times N$ matrix
$\phi:=(\phi_1,...,\phi_N)$, and the mass equations \beq
(\sigma-\tilde{m}_i{\bf 1}_k)\phi_i
=(\sigma^{\dag}-\tilde{m}_i^*{\bf 1}_k)\phi_i=0. \eeq
The D-term equation requires that $\phi$ is either zero or of rank
$k$. We shall say that the masses $\tilde{m}_i$ are {\it generic}
if there is no Higgs branch. Under this definition, equal non-zero
masses $\tilde{m}_1=\cdots=\tilde{m}_N=:\tilde{m}$ are generic. To
see this note that $(\sigma-\tilde{m}{\bf 1}_k)$ is always
non-zero because $\sigma$ is traceless. This means that the mass
equation $(\sigma-\tilde{m}{\bf 1}_k)\phi=0$ requires the rank of
$\phi$ to be less than $k$, and hence it must be zero. The masses
are non-generic if and only if there are distinct $i_1,...,i_k$
such that $\tilde{m}_{i_1}+\cdots+\tilde{m}_{i_k}=0$; in that case
there is a Higgs branch at $\sigma={\rm
diag}(\tilde{m}_{i_1},..,\tilde{m}_{i_k})$.

For generic twisted masses, the only possible flat direction is
the Coulomb branch. Let us compute the effective action in the
weakly coupled regime. We assume that $\sigma$ is diagonalizable
(\ref{diago}), with $\sigma_1+\cdots+\sigma_k=0$, and furthermore
\beq \sigma_a\ne\sigma_b \ \ \mbox{if $a\ne b$,} \label{weak} \eeq
to suppress the strong non-Abelian gauge interactions, and
\beq \sigma_a\ne \tilde{m}_i,\ \ \forall a,i, \label{heavy} \eeq
so that there are no massless charged fields. Then we can
integrate out all massive fields, leaving only the gauge
multiplets for the unbroken $U(1)^k$, to obtain the effective
superpotential:
\beq \tilde{W}=-\sum_{i=1}^N\sum_{a=1}^k
(\Sigma_a-\tilde{m}_i)\Bigl(\log(\Sigma_a-\tilde{m}_i)-1 \Bigr).
\label{theeffW} \eeq
For an $SU(k)$ theory, the $\Sigma_a$'s sum to zero, but can be
treated independently provided we add
 the Lagrange multiplier term $-\lambda\sum_{a=1}^k\Sigma_a$.
The vacuum equations are
\beqa &&\prod_{i=1}^N(\Sigma_a-\tilde{m}_i)=\e^{-\lambda},\ \
a=1,...,k,
\label{vaceq1}\\
&&\sum_{a=1}^k\Sigma_a=0.
\label{vaceq2}
\eeqa
We first show that the Coulomb branch is lifted for generic
masses, meaning that there are no continuous families of solutions
to these equations. To see this, note that such any flat direction
must allow a solution for all values of $\e^{-\lambda}$. In
particular there must be a solution with $\e^{-\lambda}=0$. Such a
solution exists if, for each $a$, there is some $i_a$ such that
$\sigma_a=m_{i_a}$. The $i_a$'s can be taken
distinct from one another
as long as $\sigma_a$'s are generically distinct
on the assumed family of solutions. Since $\sigma$ is traceless we have
$\sum_{a=1}^k\tilde{m}_{i_a}=0$. But this violates our genericity
condition on the masses. Thus, the Coulomb branch is lifted for
generic masses. We would now like to count the number of
solutions.

To simplify the analysis, we consider a particular example of
generic masses --- the case of equal masses,
$\tilde{m}_1=\cdots=\tilde{m}_N=:\tilde{m}$. The equations are
solved by
\beqa &&
\Sigma_a-\tilde{m}=\omega^{n_a}\e^{-\lambda/N},\ \ a=1,...,k,\\
&&-k\tilde{m}=(\omega^{n_1}+\cdots+\omega^{n_k})\e^{-\lambda/N},
\eeqa
where $\omega$ is $\e^{2\pi i/N}$ and $n_a$ are integers defined
modulo $N$. The second equation requires that
$Z:=\omega^{n_1}+\cdots+\omega^{n_k}$ is non-zero, in which case
we can write
\beq \Sigma_a=\tilde{m}-{\omega^{n_a}\over Z}k\tilde{m}.
\label{thevacuum} \eeq
These solutions have the property that they are unchanged by
shifts $n_a\to n_a+1$ $\forall\ a$. The condition (\ref{heavy}) is
always satisfied, while the other (\ref{weak}) is obeyed if and
only if the $n_a$'s are all distinct. Moreover, a permutation of
$n_a$'s is a gauge symmetry. Counting the number of such
$\{n_a\}$'s, modulo gauge equivalence, is exactly the same
combinatoric problem which we encountered in
Section~\ref{subsec:na}, with the answer denoted as $n(k,N)$. This
tempts us to claim that the Witten index is
\beq w(k,N)=n(k,N). \label{result} \eeq

The above analysis has a weakness: we neglected regions that
violate (\ref{weak}) or (\ref{heavy}) in which we do not have
convenient weakly coupled variables. It is possible that we may
have missed some ground states that are lurking in these regions.
In fact, for generic masses, the vacuum equations
(\ref{vaceq1})-(\ref{vaceq2}) have no solution that violates
(\ref{heavy}) but obeys (\ref{weak}). In particular,  the
potential grows towards the loci $\sigma_a=\tilde{m}_i$, making it
unlikely that a ground state is supported there. But we still have
to worry about the loci $\sigma_a=\sigma_b$ with strong
non-Abelian gauge interactions. In the next subsection, we will
prove that there are no additional contributions to the Witten
index and the result (\ref{result}) is indeed correct.

\subsection{The Absence of Contributions from Strong Coupling Regimes}
\label{subsec:embed}

To aid our search for potential vacuum states, it will prove
useful to embed the $SU(k)$ theory into a theory in which the
index is determined decisively. A natural choice is the $U(k)$
gauge theory of the type discussed in Section~\ref{sec:split}. We
introduce $N$ massless fundamental chiral multiplets $\Phi_i$ and
$N$ chiral multiplets $P^i$ in the $\det^{-1}$ representation. We
endow the $P_i$ fields with twisted masses $\tilde{m}_i^\prime$.
The scalar potential of this theory is
\bea V&=&{1\over 2e^2}\Tr|[\sigma,\sigma^\dagger]|^2
+\frac{e^2}{2}\Tr\,\left(\sum_{i=1}^N \left(\phi_i\phi_{i}^\dagger
- |p_i|^2{\bf 1}_k \right)-r{\bf 1}_k\right)^2 \nn\\ \ \ \ &&
+\ft12\sum_{i=1}^N\phi^\dagger_i\{\sigma^\dagger,\sigma\}\phi_i +
\sum_{i=1}^N\left|\Tr(\sigma)-\tilde{m}^\prime_i\right|^2|p^i|^2
\eea
We will shortly see that in the regime $r\ll 0$, this reduces to
the $SU(k)$ theory of interest. But first we examine the
opposite limit with $r \gg 0$. Here the vacuum manifold is
$\sigma=p_i=0$, while the $\phi_i$ parameterize the Grassmannian
$G(k,N)$. Undoubtedly the Witten index of this theory is
equal to the Euler
character \cite{index}, which is
$\chi(G(N,k))= {N\choose k}=N!/(N-k)!k!$.

Let us now find these vacua in the Coulomb branch analysis. Once
again placing ourselves at the weakly coupled region
\beqa
&&\sigma_a\ne\sigma_b\ \ \mbox{if $a\ne b$};\nn\\
&&\sigma_a\ne 0,\ \ \forall a;\\
&&\sum_{a=1}^k\sigma_a\ne \tilde{m}^\prime_i\ \ \forall i,\nn
\eeqa
and integrating out the heavy fields, we obtain the twisted
superpotential,
\beq \tilde{W}=-t\Sigma
-\sum_{a=1}^kN\Sigma_a[\log\Sigma_a-1]+\sum_{i=1}^N
(\Sigma-\tilde{m}^\prime_i)
\left[\log(\tilde{m}^\prime_i-\Sigma)-1\right] \eeq
where $\Sigma=\sum_{a=1}^k\Sigma_a$. The critical points of this
potential lie at
\beq \frac{\partial\tilde{W}}{\partial \Sigma_a}=0\ \ \Rightarrow\
\
\Sigma_a^N=\e^{-t^\prime}\prod_{i=1}^N(\Sigma-\tilde{m}^\prime_i)
\label{newsol}\eeq
where $t^\prime=t-N\pi i$. For non-zero twisted masses, we may
find the full set of ${N\choose k}$ isolated vacua in a
trustworthy regime. The solutions to \eqn{newsol} fall into two
categories:

\begin{itemize}
\item The first class of solution to \eqn{newsol} is analogous to
those found in Section~\ref{subsec:na}:
\beq \Sigma_a=\frac{\omega^{n_a}}{Z}\,\Sigma\ \ \ \ \ {\rm where}\
\ \ \ \
\Sigma^N=Z^N\e^{-t^\prime}\,\prod_{i=1}^N(\Sigma-\tilde{m}^\prime_i)
\label{sol1}\eeq
with $Z=\sum_a\omega^{n_a}$. Once again, we must choose distinct
integers $n_a\in\Z/N\Z$, modulo shifts $n_a\rightarrow
n_a+1$ and subject to the constraint $Z\neq 0$. There are $n(k,N)$
such choices but, this time, each gives rise to $N$ different
vacua arising from solving the $N^{\rm th}$ order polynomial equation
for
$\Sigma$. Thus this class of solutions gives us $N\,n(k,N)$ vacua.
Since dividing out by the shifts is compensated in the counting by
multiplying by $N$, the number of vacua $N\,n(k,N)$ is simply
equal to the number of distinct choices of $n_a$ such that $Z\neq
0$.

\item The second class of solutions to \eqn{sigsol} arises from
choosing integers $n_a$ such that $Z=0$. We set
\beq \Sigma_a=\omega^{n_a}S\ \ \ \ {\rm where}\ \ \ \
S^N=\e^{-t}\prod_i\tilde{m}'_i \label{sol2}\eeq
Now we have no reason to divide out by the shifts $n_a\rightarrow
n_a+1$. Thus the number of vacua in this class is the number of
distinct choices of $n_a$ such that $Z=0$.
\end{itemize}
Between the two classes of solutions, the total number of vacua is
the number of distinct choices of $n_a$, or ${N\choose k}$.
Happily, we have found all vacua in the weakly coupled regime
where $U(k)\rightarrow U(1)^k$.

So much for $r\gg 0$. Now let us look at the theory with $r\ll 0$.
Classically we have $N$ vacua labelled by $i=1,\ldots,N$ with,
\beq \phi_j=0\ ,\ \ \ \ |p^j|^2=|r|\delta^{ij}\ ,\ \ \ \ \
\Tr(\sigma)=\tilde{m}^\prime_i \label{vacua}\eeq
In each of these vacua
the $U(k)$ gauge group is broken down to
$SU(k)$ at the scale $|r|e^2$. Below this scale, each vacuum
contains
\bea \mbox{$SU(k)$ gauge theory with $N$ fundamental chiral
multiplets.} \nn\eea
In the $i^{\rm th}$ vacuum, all the fundamental chiral multiplets
inherit a common twisted mass, $\tilde{m}_j=-\tilde{m}'_i/k$
($\forall\ j$), and this sector has a well defined index $w(k,N)$.
There are also $(N-1)$ singlet fields $p^j$ with non-zero masses
$\tilde{m}_j'-\tilde{m}^\prime_i$, but they have one ground state
each and do not change the index. Since we have $N$ isolated
classical vacua, the total number of vacua in this semi-classical
analysis is $N\,w(k,N)$. But clearly this can't be all of them
since ${N\choose k}$ is not always divisible by $N$. Where are the
vacua that we've missed?!

To aid our search for the missing vacua, let's return to the
Coulomb branch analysis. The semi-classical limit is trustworthy
for $r\to-\infty$,
and we may follow the fate of each vacuum in this limit. The
solutions of the first type \eqn{sol1} survive with
$\Sigma\rightarrow \tilde{m}'_i$. These are identified with the
classical vacua \eqn{vacua}. However, the solutions of the second
type \eqn{sol2} go to infinity in the field space
$\Sigma_a\rightarrow \infty$ in the classical limit $t^\prime
\rightarrow -\infty$; these are the vacua missed by the
semi-classical discussion above. Since we found that the number of
vacua of the first type is $N\,n(k,N)$, we conclude that the
Witten index of ${\cal N}=(2,2)$ $SU(k)$ SCQD with $N$ chiral
multiplets in the fundamental representation is
indeed
\beq w(k,N)=n(k,N).
\label{wittprov}\eeq
The answer using this more careful method agrees with the direct
counting in $SU(k)$ SQCD (\ref{result}), proving that there are
indeed no contributions from the strongly coupled regime
$\sigma_a=\sigma_b$. Note also that the result \eqn{wittprov}
holds for non-equal twisted masses, $\tilde{m}_i\ne \tilde{m_j}$,
as long as they are generic.

We remind the reader that $n(k,N)$ arises from the combinatoric
problem of picking $k$ distinct mod-$N$ integers $n_a\in\Z/N\Z$,
modulo shifts $n_a\rightarrow n_a+1$, such that $Z=\sum_{a}
\e^{2\pi i n_a/N}\neq 0$. Here we briefly sketch how one may
perform this counting. Using the shift symmetry $n_a\to n_a+1$, we
may set $n_1=0$, leaving us to find a set of $(k-1)$ distinct
non-zero mod-$N$ integers $\{n_2,...,n_k\}$. If the $Z\neq 0$
condition is ignored, there are ${N-1\choose k-1}$ such sets.
However, these sets sit in families which must be identified by
the shift symmetry. For example, the $(0,n_2,...,n_k)$ must be
identified with $(-n_2,0,n_3-n_2,...,n_k-n_2)$ and so on. This
means our sets of $(k-1)$ integers are also identified:
$\{n_2,...,n_k\}$, $\{-n_2,n_3-n_2,...,n_k-n_2\}$,
...,$\{-n_k,n_2-n_k,...,n_{k-1}-n_k\}$ all lie in the same family
and collectively count only one towards $w(k,N)$. These $k$ sets
are all different if the $Z=1+\omega^{n_2}+\cdots+\omega^{n_k} \ne
0$ condition is met: For example, the replacement of the first set
by the second one is achieved by the rescaling
$\omega^{n_i}\to\omega^{n_i-n_2}$ but that would transform $Z$
into $Z\omega^{-n_2}$ which is different from $Z$ if $Z\ne 0$.
 Thus we find
\beq
n(k,N)={1\over k}\left\{{N-1\choose k-1}-\#\right\}
\label{nkN}
\eeq
where $\#$ is the number of $\{n_2,...,n_k\}$'s such that
$Z=0$.

Let us illustrate this counting for low rank gauge groups. For
$SU(2)$ with $N$ odd, there is no $n_2$ such that
$1+\omega^{n_2}=0$, and therefore $w(2,N)={1\over 2}{N-1\choose 1}
={(N-1)\over 2}$. In contrast, if  $N$ is even, only $n_2={N\over
2}$ yields $1+\omega^{n_2}=0$. Thus $w(2,N)={1\over
2}\{{N-1\choose 1}-1\}={(N-2)\over 2}$. For $SU(3)$, $Z=0$ is
impossible if $N$ is not divisible by $3$, and thus
$w(3,N)={1\over 3}{N-1\choose 2}={(N-1)(N-2)\over 6}$ in that
case. If $N$ is a multiple of $3$, only $\{n_2,n_3\}=\{{N\over
3},{2N\over 3}\}$ yields $Z=0$. Hence $w(3,N)={1\over
3}\{{N-1\choose 2}-1\}={N(N-3)\over 6}$. For higher $k$, there are
more complex patterns with $Z=0$.
%For the computation, it is important to know the number $\#$
%of $\{n_2,...,n_k\}$'s such that $Z=0$. If $N$ is divisible by $k$,
%the set  $\left\{{N\over k},{2N\over k},...,{(k-1)N\over k}\right\}$
%yields $Z=0$ since the sum of all $k$-th roots of unity vanishes,
%\beq
%\sum_{l=0}^{k-1}\omega^{lN/k}=\sum_{l=0}^{k-1}\e^{{2\pi i\over k}l}=0.
%\eeq
%For $k>3$, there may be more complicated patterns with $Z=0$.
%Suppose $k$ can be written as a sum
%$k=k_1+k_2+\cdots+k_M$ with $k_{\mu}$ ($\mu=1,..,M$)
%are integers larger than $1$ that divide $N$. Then we have
%\beq
%\sum_{\mu=1}^M\omega_{\mu}\sum_{l_{\mu}=0}^{k_{\mu}-1}
%\e^{{2\pi i\over k_{\mu}}l_{\mu}}=0
%\eeq
%and one may be able to choose $\omega_{\mu}$'s so that
%$\omega_{\mu}\e^{{2\pi i\over k_{\mu}}l_{\mu}}$ are all distinct.
%For example, let us consider the case $(k,N)=(4,6)$ which allows
%$k=4=2+2$ with $N/2=3$. Then $\vec{n}=(0,3,1,4)$ is an example with
%$Z=(1-1)+\e^{2\pi i/6}(1-1)=0$. In fact there is only one other
%example $\vec{n}=(0,3,2,5)$ and hence $\#=2$. This way we find
%$n(4,6)={1\over 4}\{{5\choose 3}-2\}=2$.
The results for low values of $k$ and $N$ are shown in Table 1 in
the Introduction.

\subsection{Varying the Twisted Masses}
\label{subsec:limits}

It is natural to claim that the index \eqn{result} truly counts
the number of supersymmetric ground states (rather than merely
counting ground states weighted with $(-1)^F$). This would mean
that the only ground states lie in the weakly coupled regime and
are solutions to (\ref{vaceq1})-(\ref{vaceq2}). We can make some
consistency checks of this proposal by following the fate of the
ground states in various limits of the twisted masses.

\noindent
\underline{$\tilde{m}\to\infty$}

We claim that $SU(k)$ with $N\leq k$ massive flavors has no
supersymmetric ground state, while the theory with $N=k+1$ has a
unique supersymmetric ground state. At first blush this is
reminiscent of ${\cal N}=1$ SQCD in four dimensions where the
$SU(k)$ theory with $N$ massless flavors has no supersymmetric
ground state for $1\leq N \leq k-1$ \cite{ADS}. However, there's
an important difference. In the four dimensional theory, if the
$N<k$ flavors have mass $m\gg \Lambda$ then supersymmetric vacua
do exist, sitting at a distance $\sim 1/m$ in field space. As
$m\rightarrow 0$, the vacua move to infinity, while as
$m\rightarrow\infty$ they coalesce around the strong coupling
scale $\Lambda$. This behavior is necessary to accommodate the
well-known fact that the Witten index for four dimensional pure
$SU(k)$ theory is $k$. In contrast, the two dimensional theories
have no vacua for $N\leq k$, even in the case of massive chiral
multiplets. Relatedly, for $N>k$ multiplets with equal masses
$\tilde{m}$, the vacua sit at a distance $\sim \tilde{m}$ in the
field space. See Eqn~(\ref{thevacuum}). If we decouple the matter
fields by sending the common twisted mass $\tilde{m}$ to infinity,
all these vacua disappear from the finite region in the field
space. Although our calculation above was only valid for theories
with $N>0$ chiral multiplets, these facts strongly suggest that
$2d$ $(2,2)$ pure $SU(k)$ Yang-Mills theory has no supersymmetric
ground state. Recent advances in constructing lattice models for
supersymmetric gauge theories (see \cite{joel} for a reveiw and
references) suggest that it may be possible to
test this claim numerically.

The situation also bears a resemblance to three-dimensional
${\mathcal N}=2$ SQCD with {\it real masses} $m_r$ which are the
$3d$ analogs of the $2d$ twisted masses \cite{Zheng}. In that
case, there is a moduli space of supersymmetric vacua consisting
of branches located at distances $\sim m_r$ from the origin of the
Coulomb branch \cite{dHO,Ofer}. As $m_r\to \infty$, the
supersymmetric branches move away to infinity and we are left with
the Coulomb branch of pure Yang-Mills theory, which is lifted by a
superpotential \cite{AHW}. In contrast, in two-dimensions the
superpotential vanishes in the $\tilde{m}\to\infty$ limit.
Nevertheless, the claimed absence of vacua in the strongly coupled
regime in $N>0$ theories suggests that in two dimensions, as in
three, there is no supersymmetric ground state in the pure ${\cal
N}=(2,2)$ Yang-Mills theory.

\noindent \underline{An Example}

For illustration let us consider the $SU(2)$ theory with $N=5$
flavors and distinct twisted masses.  We choose one chiral
multiplet to have mass $\tilde{m}_1$, while the four remaining
multiplets have mass $\tilde{m}_2$. The vacuum equations
(\ref{vaceq1})-(\ref{vaceq2}) yield the equation
\beq (\Sigma-\tilde{m}_1)(\Sigma-\tilde{m}_2)^4
=(-\Sigma-\tilde{m}_1)(-\Sigma-\tilde{m}_2)^4, \eeq for
$\Sigma:=\Sigma_1=-\Sigma_2$, which has five solutions \beq
\Sigma=\ 0,\ \pm i\tilde{m}_2,\ \pm
\sqrt{-4\tilde{m}_2^2-5\tilde{m}_1\tilde{m}_2}. \eeq
The solution $\Sigma=0$ is in the strong coupling regime
$\Sigma_1=\Sigma_2$ and must be omitted. Further, the solutions
$\Sigma=+M$ and $\Sigma=-M$ should be identified since they are
related by the permutation $\Sigma_1\leftrightarrow\Sigma_2$. Thus
there are in fact two vacua: one with $\Sigma=\pm i\tilde{m}_2$
and another with $\Sigma= \pm
\sqrt{-4\tilde{m}_2^2-5\tilde{m}_1\tilde{m}_2}$. This reproduces
the counting $w(2,5)=2$. We can now consider various limits.
Sending $\tilde{m}_2\to\infty$ decouples four of the matter fields
and, from the index $w(2,1)=0$, we expect to find no surviving
ground states. Indeed, it is simple to see that both vacua run to
infinity. Alternatively, sending $\tilde{m}_1\to \infty$ decouples
only a single chiral multiplet. Now the vacuum at $\Sigma= \pm
\sqrt{-4\tilde{m}_2^2-5\tilde{m}_1\tilde{m}_2}$ runs away, while
the vacuum at $\Sigma=\pm i\tilde{m}_2$ remains. This is in
agreement with the counting $w(2,4)=1$.

\medskip
\noindent
\underline{$\tilde{m}\to 0$}

The claim that there are no supersymmetric ground states for
$N\leq k$ massive flavors implies that there are also no
normalizable supersymmetric ground states when the twisted masses
are turned off. To see this, suppose that it was not true and a
normalizable supersymmetric ground state appeared in the massless
theory. Then turning on masses would only improve the
long-distance behavior of the wavefunction and the ground state
would survive in the massive theory as well, in contradiction to
our claim. We will revisit massless SQCD in Section~\ref{sec:IR}.

\subsection{Complex Masses for $SU(2)$ Theories}
\label{subsec:cplxmass}

We have seen that as the twisted masses $\tilde{m}\rightarrow
\infty$, they carry some vacua to infinity with them in the
$\sigma_a$ field space. In this manner, ground states decouple
from the theory as the number of chiral multiplets decrease.

For $SU(2)$ SQCD with $N$ fundamentals, the theory admits another,
complex, mass term. This is allowed because the baryons are
quadratic and one may add a gauge invariant superpotential,
\beq W=\sum_{i,j=1}^Nm^{ij}\epsilon_{ab}\Phi^a_i\Phi^b_j.
\label{cmass} \eeq
(There is no analog of this in $SU(k)$ SQCD with $k>2$ since the
gauge invariants have power $k$ or higher in the fundamentals
$\Phi_i$.) We would like to understand the effect of these complex
masses on the ground state spectrum. By continuity, any
supersymmetric vacua that exist at $m^{ij}=0$ must survive at
finite $m^{ij}$, and the index remains $n(2,N)=[{N-1\over 2}]$.
The question we want to ask is: what happens as some part of
$m^{ij}$ is sent to infinity? Suppose a rank $2l$ sub-matrix
becomes infinite, decreasing the number of flavors from $N$ to
${N-2l}$. The number of vacua must correspondingly decrease by
$l$. But how does this happen? By the decoupling of the chiral and
the twisted chiral sectors \cite{book}, $m^{ij}$ cannot enter into
the twisted superpotential. This rules out the possibility that
the locations of the vacua move to infinity in the $\sigma_a$
space as we vary the masses $m^{ij}$. There must be a different
mechanism at work.

The key to understanding the fate of the vacua is the derivation
of the correct effective Lagrangian on the Coulomb branch in the
presence of the complex mass. Let us study this in a  simpler
model: $U(1)$ gauge theory with two fields $\Phi_1$, $\Phi_2$ of
charge $1$, $-1$, each with a common twisted mass $\tilde{m}$,
together with the complex mass term
\beq W=m\Phi_1\Phi_2. \eeq
The effective Lagrangian is obtained by integrating out $\Phi_1$
and $\Phi_2$. The bosonic determinants yield the potential term
\beq \int{\dd^2 k\over (2\pi)^2}\Bigl[
\log(k^2+|\sigma-\tilde{m}|^2+|m|^2+D)
-\log(k^2+|\sigma+\tilde{m}|^2+|m|^2-D)\Bigr] \eeq
for a constant profile of $\sigma$ and $D$. If $|m|\ll
|\sigma\pm\tilde{m}|$, one may expand the log as
\beq \log(k^2+|\sigma\mp \tilde{m}|^2+|m|^2\pm D)
=\log(k^2+|\sigma\mp\tilde{m}|^2)+ {|m|^2\pm D\over
k^2+|\sigma\mp\tilde{m}|^2}+\cdots. \eeq
The first term is cancelled by the fermionic determinants while
the second term leads to a part of the twisted F-term associated
with the twisted superpotential
\beq \tilde{W}
=-t\Sigma-(\Sigma-\tilde{m}_1)\Bigl[\log(\Sigma-\tilde{m}_1)-1\Bigr]
+(\Sigma+\tilde{m}_2)\Bigl[\log(-\Sigma-\tilde{m}_2)-1\Bigr].
\label{twtw} \eeq
We have employed twisted superpotentials of this type throughout
this paper. On the other hand, in the opposite regime
$|m|\gg|\sigma\pm\tilde{m}|$, the above expansion is not valid and
it is more appropriate to write
 \beq \log(k^2+|\sigma\mp
\tilde{m}|^2+|m|^2\pm D) =\log(k^2+|m|^2)+ {|\sigma\mp
\tilde{m}|^2\pm D\over k^2+|m|^2}+\cdots. \label{dunno}\eeq
We do not have the twisted superpotential (\ref{twtw}) but some
other effective Lagrangian in terms of the same field $\Sigma$.
While we do not specify precisely this Lagrangian, it is
sufficient to note that it can be expanded in powers of
$(\sigma\pm \tilde{m})/m$ and vanishes in the limit
$|m|\to\infty$.

\begin{figure}[htb]
\centerline{\includegraphics{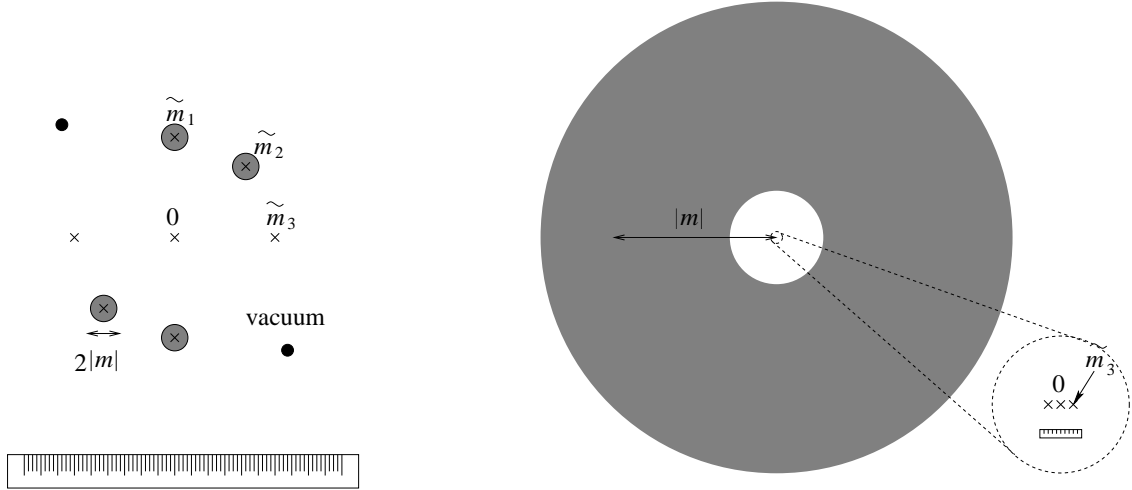}}
\caption{The Coulomb branch for $SU(2)$ SQCD with $N=3$.
The scales of the left part ($|m|\ll |\tilde{m}_i|$)
and the right part ($|m|\gg |\tilde{m}_i|$) are different.}
\label{pushed}
\end{figure}

We now apply this consideration to our $SU(2)$ SQCD. Let us first
consider a specific example: $N=3$ flavors with generic twisted
masses $\tilde{m}_1$, $\tilde{m}_2$, $\tilde{m}_3$, and with a
complex mass $m$ for the first and the second fields
\beq W=m\epsilon_{ab}\Phi^a_1\Phi^b_2. \eeq
If $|m|\ll |\tilde{m}_i|$, the effective theory based on the
superpotential (\ref{theeffW}) is valid except inside the two
discs with radius $|m|$ and centers $\sigma_1=\pm\tilde{m}_1$,
$\pm\tilde{m}_2$. See the left part of Figure~\ref{pushed}.
The one vacuum state localized at the critical point of
(\ref{theeffW}) is essentially unaffected by the presence of these
discs. However, as $|m|$ is increased past $|\tilde{m}_i|$'s, the
discs grow, merge into one domain and swallow the critical point.
Inside that domain, the effective theory has the F-term potential
associated with only the $i=3$ part of the superpotential
\beq
\tilde{W}_3=-(\Sigma_1-\tilde{m}_3)(\log(\Sigma_1-\tilde{m}_3)-1)
+(\Sigma_1+\tilde{m}_3)(\log(-\Sigma_1-\tilde{m}_3)-1), \eeq
plus some additional Lagrangian ${\mathcal L}_{12}$. Even when
$|m|\gg |\tilde{m}_i|$, the ground state must remain as long as
$|m|$ is finite, since the Lagrangian based on the full
superpotential (\ref{theeffW}) is still valid for $|\sigma|\gg
|m|$, ensuring that no state has run away to infinity. The ground
state is therefore supported within the domain $|\sigma| < |m|$.
Deep inside the domain, the additional Lagrangian ${\mathcal
L}_{12}$ is very small and the potential arising from  $\tilde{W}_3$
dominates. But, as we have seen in Sections~\ref{subsec:count} and
\ref{subsec:embed}, this potential does not have a zero and pushes
the ground state wavefunction away from the center of the Coulomb
branch. We conclude that the ground state can have support only in
a halo region of radius $\sim |m|$, as depicted in the right part
of Figure~\ref{pushed}. As $|m|$ is sent to infinity, the halo
disappears from our sight. This is how the unique ground state of
the $N=3$ theory disappears as the complex mass is sent to
infinity.

To end this discussion, let us add further flavors with twisted
masses $\tilde{m}_4,...,\tilde{m}_N$, but without complex masses.
If $|m|\ll |\tilde{m}_i|$, there are $[{N-1\over 2}]$ ground
states localized at the critical points of (\ref{theeffW}). If
$|m|\gg |\tilde{m}_i|$, there are as many ground state as before
but they are no longer localized at the same critical points. The
effective Lagrangian inside the domain of radius $|m|$ is the
F-term associated with the $i=3,...,N$ part of the superpotential,
$\tilde{W}_{3...N}$, together with another Lagrangian arising from
the expansion of the type \eqn{dunno}. $[{N-1\over 2}]-1$ states are localized
at the critical points of $W_{3...N}$ but one state is not. As we
will explain in more detail in Section~\ref{subsec:criterion},
the character of the
F-term potential in the region $|\tilde{m}_i|\ll |\sigma|\ll |m|$
depends on whether $N$ is even or odd. For $N$ even,  the F-term
potential is nearly zero while, for $N$ odd, it approaches a
positive constant density $e^2\pi^2/2$ (see
Section~\ref{subsec:criterion}). Therefore we conclude that, if
$N$ is odd, the positive potential forces the extra state to lie
in a halo of radius $\sim |m|$. In contrast, if $N$ is even, the
state may have extended support throughout the region
$|\tilde{m}_i|\ll |\sigma| <|m|$. As $|m|$ is sent to infinity,
for odd $N$ the extra state disappears as the halo moves away to
infinity. For even $N$ the state spreads, and become
non-normalizable. In both cases, we are left with the $[{N-1\over
2}]-1$ localized states.

\section{Infra-Red Dynamics of $SU(k)$ Gauge Theories}
\label{sec:IR}

In this section, we study $SU(k)$ gauge theories with $N$ {\it
massless} fundamental chiral multiplets. This class of theories
includes massless SQCD, as well as theories with superpotentials.
We expect that such a theory flows to a non-trivial superconformal
field theory (SCFT) in the infra-red limit. However, the existence
of the Coulomb branch presents a potential problem for the
infra-red dynamics. The Coulomb branch provides a flat non-compact
direction in field space into which low-energy states may spread.
Typically, the existence of such a Coulomb branch is signalled by
divergences in the correlation functions of Higgs branch
operators. Such behavior is seen at the conifold points of ${\cal
N}=(2,2)$ Abelian theories \cite{phases,comments} and at the
special point of ${\cal N}=(4,4)$ SQED \cite{comments,Higgs,AB}.
In this section, we will derive the criterion for the existence of
a quantum Coulomb branch and analyze the associated singularity.
We start with the computation of the central charge of the SCFT.

\subsection{IR Central Charge}

In an ${\cal N}=(2,2)$ supersymmetric gauge theory with a simple
gauge group, the axial $U(1)$ R-symmetry is anomaly free because
$\Tr F_{01}=0$ for any representation of the group. If the
superpotential is homogeneous and respects the vector $U(1)$
R-symmetry, the theory is expected to flow to an ${\cal N}=(2,2)$
SCFT with the left and right $U(1)$ current algebras inherited
from the vector and axial R-symmetries.

In such a case, one can learn much about the CFT by studying the
chiral ring \cite{LVW} of the UV gauge theory. In particular, the
$\oQ_+$-chiral ring includes a copy of the ${\mathcal N}=2$
superconformal algebra that descends to become part of the
right-half of the chiral algebra in the infra-red theory
\cite{minimal}. It consists of the R-current $j_-$ and its
superpartners $G_-,\overline{G}_-,T_-$ which combine into a
super-R-current ${\mathcal J}$ obeying $\overline{D}_+{\mathcal
J}=0$, The central charge $\whc=c/3$ of the CFT can be read from
the $j_-$-$j_-$ operator product expansion
\beq j_-(x)j_-(y)\sim -{\whc\over (x^--y^-)^2}. \label{OPE} \eeq
This computation has been performed in several examples
--- Landau-Ginzburg models \cite{minimal}, ${\cal N}=(0,2)$ Abelian linear
sigma models (where only $T_-$-$T_-$ OPE matters \cite{SilW}), and
${\cal N}=(2,2)$ Abelian linear sigma models \cite{hk}. However,
provided the left and right R-symmetries are correctly identified
in the UV, there is a quicker way to compute the central charge.
The idea is to couple the vector $U(1)$ R-symmetry to a gauge
field and look at the anomaly that arises for the axial $U(1)$
R-symmetry. The axial anomaly takes the form
\beq
\partial_+j_-=a F_{+-},
\label{anomaly}
\eeq
and $\partial_-j_+=aF_{+-}$,
where $F_{+-}=\partial_+A_--\partial_-A_+$ is
the curvature of the $U(1)_V$ gauge field.
Taking the variation of a correlator including
(\ref{anomaly}) with respect to
$A_+\to A_++\delta A_+$ and setting $A=0$, we find
\beq \left\langle \partial_+j_-(x)\ {i\over\pi}\int\delta
A_+(y)j_-(y)\,\dd^2y\,\, O_1\cdots O_s \right\rangle
=-a\,\partial_-\delta A_+(x)\Biggl\langle O_1\cdots O_s
\Biggr\rangle. \eeq
It follows that the current-current product expansion is of the
form (\ref{OPE}) with \beq \whc=a. \eeq Thus, the infra-red
central charge can be read by computing the axial anomaly in the
system where the vector R-symmetry is gauged. In a system with
Dirac fermions $\psi_{i\pm},\opsi_{i\pm}$ where $\psi_{i\pm}$ has
$U(1)_V$ charge $q_i$ and $U(1)_A$ charge $\mp 1$, the number $a$
that measures the anomaly is minus the sum of the $U(1)_V$
charges;
\beq a=\sum_{i:\,\, {\rm Dirac\, \,fermion}}(-q_i).
\label{formulaa} \eeq
Let us apply this technique to the $SU(k)$ gauge theory with $N$
massless fundamental chiral multiplets and a superpotential which
is  homogeneous of degree $d$ in the baryonic variables
$B_{i_1...i_k}$,
\beq W=G_d(B) \eeq
Since the superpotential must have vector and axial R-charges $2$
and $0$, the R-charges of $B_{i_1...i_k}$ are $2/d$ and $0$,
implying that the constituent fundamental chiral multiplets
$\Phi^a_i$ have R-charges $2/dk$ and $0$. In particular, the
fermionic components $\psi^a_{i\pm}$ have vector R-charge
$(2/dk-1)$ and axial R-charge $\mp 1$. The gaugino $\lambda_{\pm}$
has vector and axial R-charges $1$ and $\mp 1$ respectively.
Adding these gives the central charge in the infra-red limit,
\beq \whc=Nk\times \left(-{2\over dk}+1\right)+(k^2-1)\times (-1)
={N(dk-2)\over d}-(k^2-1). \label{ckNd} \eeq
As a check, let us consider the case $k=N=1$: a theory with no
gauge group $SU(1)=\{1\}$, one variable $B=\Phi$ and degree $d$
superpotential $W=\Phi^d$. This is a Landau-Ginzburg (LG) model
that flows to the ${\mathcal N}=(2,2)$ minimal model of level
$(d-2)$ in the infra-red limit \cite{Martinec,VW}. The formula
(\ref{ckNd}) yields $\whc=(d-2)/d=1-2/d$, which is indeed the
central charge for the minimal model (see for example \cite{ZF}).

\subsection{The Singularity Criterion}
\label{subsec:criterion}

Let us now examine the potential singularity of the infra-red
theory. To be specific, we first consider  $SU(k)$ SQCD, the
theory without superpotential. As mentioned earlier, the classical
theory has both a non-compact Higgs branch of complex dimension
$k(N-k)+1$ and a non-compact Coulomb branch of dimension $k-1$.
How do quantum effects change this story? The Higgs branch
survives, and the sigma model on it flows to an interacting SCFT.
It has an asymptotic region where the sigma model is weakly
coupled, which implies that the central charge is the dimension
$\whc=k(N-k)+1$. In fact, in that region the R-symmetry cannot act
non-trivially on the bosonic coordinate fields \cite{comments},
which fixes the vector and axial R-charges of $\Phi^a_i$ to be
zero. The general formula (\ref{formulaa}) indeed yields
\beq \whc=Nk\times (-(-1))+(k^2-1)\times (-1)=Nk-k^2+1. \eeq
This theory itself is singular in the sense that the target space
is non-compact in the asymptotic large $\phi$  directions. But we
would like to focus on another type of singularity of the Higgs
branch theory
--- the additional non-compactness at $\phi=0$
associated with the existence of Coulomb branch \cite{comments}.

We may repeat the analysis of the effective superpotential
\eqn{tildew} in our model, now with a Lagrange multiplier required
to ensure that $\Sigma=\sum_{a=1}^k\Sigma_a=0$. It is not hard to
see, following the analysis of Section~\ref{subsec:na}, that most
of the Coulomb branch is lifted by quantum effects.
Coulomb branches of at most one dimension may survive, with
\beq \Sigma_a=\omega^{n_a}\,\e^{-\lambda/N}\eeq
where $\omega^N=1$ and $n_a\in \{0,1,\ldots, N-1\}$. This is a
suitable $SU(k)$ configuration only if the $k$ distinct integers
$n_a$ can be found such that $Z=\sum_a\omega^{n_a} =0$. 
If no such
integers $n_a$ can be found, then the Coulomb branch is completely
lifted by quantum effects.

For example, for $SU(2)$ gauge theory with $N$ fundamental chiral
multiplets, there exists a quantum Coulomb branch if
and only if $N$ is even.
To see this explicitly, we write
the effective superpotential
on the Coulomb branch obtained by integrating out
the $N$ chiral multiplets:
\beq
\tilde{W}=-N\Sigma_1(\log(\Sigma_1)-1)+N\Sigma_1(\log(-\Sigma_1)-1)
=N\pi i\Sigma_1.
\label{su2QCD}
\eeq
This shows that a theta angle $\theta=N\pi$
for the unbroken $U(1)$ subgroup is generated.
The theta angle is identified as a background electric field
\cite{coleman} that carries
the electro-static energy with density $e^2\theta^2/2$.
If $N$ is even, pair creations of the charge $1$ particles
will completely screen the background electric field:
then the energy density is zero and the Coulomb branch remains as
an exact flat direction.
If $N$ is odd, pair creations cannot completely screen it
--- the field with $\theta=\pi$ or $-\pi$ will always remain.
The Coulomb branch is lifted.
For $SU(3)$ gauge theory,
there is one
Coulomb branch if and only if $N$ is divisible by $3$,
in the direction $(\Sigma_1,\Sigma_2,\Sigma_3)
=(S,\e^{2\pi i/3}S,\e^{4\pi i/3}S)$.
For $SU(k)$ with higher $k$,
there can be multiple one-dimensional Coulomb branches.
For example, for $SU(4)$ theory with $N=8$
fundamentals, there are {\it two} Coulomb branches:
one is in the direction
 $(\Sigma_1,\Sigma_2,\Sigma_3,\Sigma_4)=(S,iS,-S,-iS)$
and another is along $(S,-S, \e^{2\pi i/8}S,-\e^{2\pi i/8}S)$.

To summarize, we propose a criterion for the existence of the
one-dimensional Coulomb branches or, equivalently, 
the criterion for a singularity
in the infra-red fixed point of the Higgs branch theory: {\it
$SU(k)$ SQCD with $N$ massless flavors is singular at the origin
$\phi=0$ if and only if there are $k$ distinct $N$-th roots of
unity whose sum vanishes.}
It would be an interesting problem to study such singular conformal
field theories, particularly when there are multiple Coulomb branches.

This criterion is applicable also in theories with a non-trivial
superpotential $W$, because the computation of the effective
superpotential $\tilde{W}$ for the twisted chiral superfield is
not affected. If $W$ removes the non-compact Higgs branch and if
the criterion reveals no singularity, then we obtain a completely
regular superconformal field theory in the infra-red limit. For
example, let us consider the superpotential $W=G_d(B)$ which is a
homogeneous polynomial of degree $d$ in the baryons
$B_{i_1...i_k}$. Suppose $G_d(B)$ is generic so that the equation
in the Pl\"ucker coordinates $G_d(B)=0$ defines a smooth
hypersurface of the Grassmannian $G(k,N)$. Then the F-term
equation
\beq {\partial W\over
\partial\Phi^a_j}
=\sum_{i_1<\cdots <i_k}{\partial G_d(B)\over \partial
B_{i_1...i_k}}\  {\partial B_{i_1...i_k}\over \partial \Phi^a_j}=0
\eeq
together with the D-term equation (\ref{Deq}) has no solution
other than $\Phi^a_i=0$. This ensures that the superpotential
removes the non-compact Higgs branch. The assignment of non-zero
R-charge to $\Phi_i^a$ is then justified and we can safely say
that the IR fixed point has central charge
$\whc=N(dk-2)/d-(k^2-1)$, as we have computed in (\ref{ckNd}). Let
us denote this fixed point by ${\mathcal C}_{k,N}(G_d)$. We claim
that {\it the superconformal field theory ${\mathcal
C}_{k,N}(G_d)$ is completely non-singular if and only if there are
no $k$ distinct $N$-th roots of unity that sum to zero.}

\subsection{$N=k+1$: Duality with Free SCFT and LG Models}
\label{subsec:confinement}

Perhaps the most striking aspect of our singularity criterion is
that for many values of $(k,N)$, the Coulomb branch is lifted and
the theory is non-singular at the origin $\phi=0$. For any $k$,
the $SU(k)$ gauge theory with $N=k+1$ fundamentals is such an
example. We would like to comment here on some consequence of the
regularity in this class of examples.

Let us first consider  SQCD, the theory without a superpotential:
$W=0$. For $N=k+1$ there are no algebraic relations among the $N$
baryonic operators
\beq B^i:=\epsilon^{ij_1...j_k}B_{j_1...j_k}, \qquad i=1,...,N.
\eeq
These provide global coordinates of the Higgs branch which is
therefore holomorphically isomorphic to $\C^N$. Geometrically,
however, the classical Higgs branch
 is a cone with  the metric

\beq \dd s^2={|\!| \dd B |\!|^2\over \,\,\ |\!| B
|\!|^{2-2/k}\!\!\!}. \label{Hmetric} \eeq
To see this, we first note that the global symmetry group $U(N)$
transforms the baryons as $B^i\mapsto B^j(g^{-1})^{\,\,i}_{j}\det
g$ and hence the Higgs branch is a cone over the sphere
$S^{2N-1}$. The metric can therefore be written as $\dd s^2=f|\!|
\dd B |\!|^2$ for some function $f=f(|\!|B|\!|)$. To fix the
function $f$, we use the fact that the Higgs branch of the $N=k$
theory embeds into that of the $N=k+1$ theory, and is identified
as, say, the complex line $B_2=\cdots= B_N=0$. The $N=k$ Higgs
branch is simply the cone $\C/\Z_k$ with deficit angle
$2(k-1)\pi/k$ and hence we find $f=1/|\!|B|\!|^{2-2/k}$.

The classical metric (\ref{Hmetric}) has a conical singularity at
the origin $B=0$, the unique enhanced $SU(k)$ symmetry point. This
singularity comes from elimination of the $SU(k)$ gauge multiplets
that becomes massless at this point.
However, we have learnt that this must be modified
by quantum effects. In the quantum theory, we have seen that the
Coulomb branch is lifted by the electrostatic energy with density
$\sim e^2\pi^2/2$, ensuring that the $SU(k)$ gauge multiplet has a
mass of order $e$. The theory at energies much smaller than $e$ is
described purely in terms of the baryons $B^1,...,B^N$ as the
independent variables, and no
singularity is expected at the origin $B=0$. We conclude that the
low energy theory is just the sigma model on the quantum Higgs
branch where the conical singularity at $B=0$ is smoothed out. We
expect that the sigma model metric simply flattens at lower
energies and that the infra-red fixed point is the sigma model on
the flat $\C^N$. In summary, {\it $SU(k)$ SQCD with $N=k+1$
massless flavors flows to a
 free superconformal field
theory with $\whc=N$, described by the $N$ baryonic operators.}

This behavior is reminiscent of four dimensional ${\mathcal N}=1$
$SU(N_c)$ SQCD with $N_f=N_c$ flavors \cite{Seiberg}, where the
singularity of the classical moduli space of vacua is smoothed by
instanton corrections. However, in the four-dimensional example,
the complex structure is modified and the moduli space is moved
away from the point of symmetry enhancement.
In contrast, in our two dimensional
example, the symmetry enhancement point
 $B=0$ remains in the moduli (or the target) space; the
metric is merely smoothly rounded there.

Let us now turn to the theories with non-trivial superpotential,
which we take to be a homogeneous polynomial of degree $d$ in the
baryonic coordinates,
\beq W=G_d(B^1,...,B^N). \label{sup1} \eeq
As in 2d SQCD, the theory at energies much smaller than $e$ is
described in terms of the gauge invariant composites
$B^1,...,B^N$. We claim that {\it the theory is dual at energies
$\ll e$ to the Landau-Ginzburg model with $N$ variables
$X^1,...,X^N$ and superpotential}
\beq W=G_d(X^1,....,X^N). \label{sup2} \eeq
{\it In particular, the two theories flow to the same infra-red fixed
point.} We recall that the fixed point of the $SU(k)$
theory, denoted as ${\mathcal C}_{N-1,N}(G_d)$, has central charge
\beq \whc=Nk\times\left(1-{2\over dk}\right)-k^2+1 =k+1-N{2\over
d}=N\left(1-{2\over d}\right). \eeq
This matches the central charge for the IR fixed point of the
Landau-Ginzburg model \cite{Martinec,VW}.
For $d=1$ and $2$, the formula yields $\whc=-N$ and $0$ respectively,
and the theories do not flow to
non-trivial fixed points in the deep infra-red limit.
However, the low energy duality with the Landau-Ginzburg model
should still hold.
We examine these cases now:

\noindent
\underline{$d=1$:}\\
[0.2cm] The superpotential is linear in baryons,
$W=a_1B^1+\cdots+a_NB^N$.
The supersymmetry
condition in the dual Landau-Ginzburg
 model reads $\partial_iW=a_i=0$ ($\forall\
i$) and has no solution. Hence, when $N=k+1$, a linear
superpotential induces supersymmetry breaking for the low energy
theory on the Higgs branch! For the $SU(2)$ case ($N=3$), the
linear superpotential means that the fundamentals have complex
masses (\ref{cmass}), $m^{ij}=a_k\epsilon^{kij}$, and one can
actually understand this supersymmetry breaking from our study in
Section~\ref{subsec:cplxmass}. We have seen there that, with a
generic twisted mass, there is one supersymmetric ground state
supported in a wide halo region of radius $\sim |\!|a|\!|$ in the
$SU(2)$ Coulomb branch. There is no supersymmetric ground state
near the center where the Higgs branch is located. This conclusion
must continue to hold when the twisted masses vanish, as we argued
in Section~\ref{subsec:limits}. Thus, this Higgs branch theory has
no supersymmetric ground state.

\noindent
\underline{$d=2$:}\\
[0.2cm] Let us choose a superpotential that is non-degenerate, say
$W=m_1(B^1)^2+\cdots +m_N(B^N)^2$ with all $m_i\ne 0$. The dual
Landau-Ginzburg model is a theory with a unique supersymmetric
ground state which has mass gap ${\rm min}\{|m_i|\}$. We conclude
that the $SU(k)$ theory with a non-degenerate quadratic
superpotential for the baryons also has a unique supersymmetric
ground state with a mass gap.

A variant of this example arises when the baryons are coupled to
singlets in a quadratic superpotential. For example, let us
consider the $SU(2)$ theory with $N=3$ fundamentals
$\Phi_1,\Phi_2,\Phi_3$ and $3$ singlets $A_1,A_2,A_3$ which are
coupled through the superpotential
\beq
W=\epsilon^{ijk}A_i\epsilon_{ab}\Phi^a_j\Phi^b_k=A_1B^1+A_2B^2+A_3B^3.
\label{massiveth} \eeq
At energies much lower than $e$, this is dual to a Landau-Ginzburg
model of six variables $X^i$, $A_i$ with superpotential
$W=A_1X^1+A_2X^2+A_3X^3$. In particular, the theory has a unique
supersymmetric ground state with a mass gap. This example will
prove important in Section~\ref{sec:Rodland}.

For the range $3\leq d \leq N$, we will provide an alternative
derivation of the duality in Section \ref{subsec:rgflow}.

\subsection{Interlude -- The $U(k)$ Linear Sigma Model Revisited}
\label{subsec:inter}

\newcommand{\wtS}{\widetilde{\Sigma}}

Let us return to the $U(k)$ linear sigma models that we introduced
in Section~\ref{sec:split}. We saw that, for $r<0$, there exist
mixed Coulomb-Higgs branches (see the discussion below
\eqn{diago}) and we postponed the analysis of the fate of these
branches until a suitable later time. That time is now.

For a negative FI parameter, the $p^{\alpha}$ fields have vevs and
the low energy gauge group is the unbroken $SU(k)$. According to
the criterion derived in Section~\ref{subsec:criterion},
this low energy theory is
singular for certain $(k,N)$ because of the existence of the
quantum Coulomb branch. Does this mean that the $U(k)$ linear
sigma model has an infra-red singularity for these $(k,N)$? Among
the models giving rise to Calabi-Yau three-folds (see Table 2),
those with $(k,N)=(2,4)$, $(2,6)$ and $(3,6)$ should be singular
by our criterion. Yet the mirror analysis of \cite{Duco} shows no
sign of a singularity in the finite $r \ll 0$ region of the moduli
space. Moreover, an explicit computation of correlation functions
in the Higgs branch theory similarly shows no hint of singular
behavior for $r\ll 0$ \cite{HT2}. What is going on?

We claim in fact that  the mixed Coulomb-Higgs branch is lifted and
the $U(k)$ theory has no singularity
for any $(k,N)$ as long as $r\ll 0$ is finite. This is due to a
residual, non-trivial effect, left behind by the Higgsed $U(1)$
sector. To confirm
this, let us compute the effective superpotential on the $SU(k)$
Coulomb branch at finite $r\ll 0$. We give
slowly varying, large and distinct, traceless background eigenvalues $\wtS_a$
to the fieldstrength $\Sigma$,
\beq \sum_{a=1}^k\wtS_a=0. \eeq
These eigenvalues set an energy scale which we write $M$. The
fundamental fields $\Phi_i$ and W-bosons all have mass of order
$M$, and must be integrated out to obtain the effective theory
below the scale $M$. This effective theory consists of the
$U(1)^k$ vector multiplets and $S$ chiral multiplets $P^{\alpha}$
that are charged only under the trace part $U(1)_0$ of $U(1)^k$.
(There is also a twisted superpotential, a remnant of the
$\Phi_i$'s.) We choose the basis of the charge lattice of $U(1)_0$
so that $P^{\alpha}$ has charge $q_{\alpha}$. The sector of
$U(1)_0$ and the $P^{\alpha}$ fields is essentially that of the
linear sigma model for the weighted projective space ${\mathbb
W}{\mathbb P}^{\,q_1,\ldots,q_s}$, with the effective FI-parameter
$r_0(M)=-r\gg 0$ at the cut-off scale $M$. The FI parameter $r_0$
runs towards smaller values as the energy is decreased and
eventually enters into the negative region. Then it is appropriate
to integrate out the $P^{\alpha}$ fields, just as we do in the
$\CP^{N-1}$ model, and we obtain an effective action for
$\Sigma_0$ with ``coupling constants'' $\wtS_a$. Since it is the
same thing as integrating out $\Phi_i$'s and $P^{\alpha}$'s at the
same time, obviously the end result is exactly equal to
(\ref{tildew}) with the replacement
 \beq
\Sigma_a=\wtS_a-\Sigma_0/k. \label{rede} \eeq
The final step is to integrate out $\Sigma_0$, i.e. extremize the
superpotential with respect to $\Sigma_0$. Solving for $\Sigma_0$
in terms of $\wtS_a$'s gives
\beq \Sigma_0=f(\e^{-t},\wtS). \label{efftw} \eeq
One may say that
the Higgsed $U(1)$ sector has dynamically induced an effective twisted mass
$\Sigma_0/k=f(\e^{-t},\wtS)/k$ for the fundamentals of the $SU(k)$
theory.
Plugging (\ref{efftw}) back into the superpotential,
we obtain a non-trivial superpotential for $\wtS_a$'s.
This
lifts the $SU(k)$ Coulomb branch at finite $r\ll 0$.

Since the last steps above are computationally involved, let us illustrate
how this works in the case of $SU(2)$. The twisted superpotential
for $\Sigma_0$ and ``coupling'' $\wtS_1=-\wtS_2=:\wtS-\Sigma_0/2$
is\footnote{Instead of the symmetric but formal ``embedding''
$\Sigma_1=\wtS-\Sigma_0/2$, $\Sigma_2=-\wtS-\Sigma_0/2$, we work
with a genuine embedding $\Sigma_1=\wtS-\Sigma_0$,
$\Sigma_2=-\wtS$ of $U(1)_0$ into $U(1)^2$. This is to respect the
integral structure of charge lattice which is important when a
pair creation of quantized charges is involved in the discussion.}
\beqa \tilde{W}
&=&t\Sigma_0-N(\wtS-\Sigma_0)(\log(\wtS-\Sigma_0)-1)
-N(-\wtS)(\log(-\wtS)-1)
\nn\\
&&+\sum_{\alpha=1}^Sq_{\alpha}\Sigma_0(\log(-q_{\alpha}\Sigma_0)-1).
\eeqa
The extremum equation $0={\partial \tilde{W}/\partial\Sigma_0}
=t+N\log(\wtS-\Sigma_0)-\sum_{\alpha=1}^Sq_{\alpha}\log(q_{\alpha}\Sigma_0)$
is solved by
\beq \Sigma_0=f(\e^{-t})\wtS, \eeq
where $f=f(\e^{-t})$ solves the equation
 \beq
\prod_{\alpha=1}^Sq_{\alpha}^{q_{\alpha}}f^N=\e^t(1-f)^N.
\label{fdef} \eeq
If we now plug this solution back into the superpotential, we find
\beq \tilde{W}=\Bigl[-N\log(1-f)+N\log(-1)\Bigr]\wtS.
\label{SU2sup} \eeq
An exact Coulomb branch exists only when the coefficient of $\wtS$
vanishes (modulo $2\pi i \Z$): that is, when $(f-1)^N=1$. This is
solved by $f=1+\omega$ where $\omega^N=1$ or when $t$ is given by
\beq
\prod_{\alpha=1}^Sq_{\alpha}^{q_{\alpha}}(1+\omega)^N=\e^t(-1)^N,
\eeq
for $\omega^N=1$. This solution lies in the trustworthy regime
only when $\omega\ne \pm 1$. But this is nothing other than the
singular point (\ref{sigsing}); we have simply recovered the
original Coulomb branch obtained in Section~\ref{sec:split}. We
find that the classical $SU(2)$ Coulomb branch at $r\ll 0$ is
indeed completely lifted, both for odd $N$ (as we saw for pure
$SU(2)$ SQCD in Section~\ref{subsec:criterion})
 and also for even $N$.

Despite the discussion above, we expect that in the strict limit
$r\rightarrow -\infty$, where the Higgs mass $e\sqrt{-r}$ goes to
infinity, the effect of the $U(1)_0$ vanishes. Indeed, this can be
seen in our calculation since, in this limit, $f(\e^{-t},\wtS)$
vanishes (see, for example, (\ref{fdef}) for the $SU(2)$ case).
The superpotential on the $SU(k)$ Coulomb branch should converge
to that of the pure $SU(k)$ gauge theory with massless flavors.
For the $SU(2)$ example above, in the limit $f\rightarrow 0$ one
indeed recovers \eqn{su2QCD} from (\ref{SU2sup}). Thus, according
to our result of Section~\ref{subsec:criterion}, a one-dimensional
Coulomb branch develops at $t=-\infty$ when there are $k$ distinct
$N$-th roots of unity that sum to zero. For such $(k,N)$, we
expect that the CFT at $t=-\infty$ is singular.

\subsection{RG Flow and Duality}
\label{subsec:rgflow}

We may extend the discussion of linear sigma models to $U(k)$
theories in which the axial $U(1)$ R-symmetry is anomalous. For
illustration, let us consider the model consisting of $N$
fundamental chiral multiplets $\Phi_1,...,\Phi_N$ and one field
$P$ in the $\det^{-d}$ representation, with superpotential
\beq W=PG_d(B), \eeq
where $G_d(B)$ is a degree $d$ polynomial in the baryons
$B_{i_1...i_k}$. At $r\gg 0$ the theory reduces to the non-linear
sigma model on the hypersurface $X_{k,N}(G_d)$ of the Grassmannian
$G(k,N)$ defined by the equation $G_d=0$. We assume the $G_d$ is
generic so that the hypersurface is non-singular. At $r\ll 0$, the
field $P$ acquires a non-zero value and breaks the gauge group
$U(k)$ to the subgroup $G$ of elements $g$ such that $\det^d g=1$,
which includes $SU(k)$ as a subgroup of index $d$, $G/SU(k)\cong
\Z_d$. In the strict $r\to -\infty$ limit, the theory has  gauge
group $G$ coupled to $N=k+1$ fundamentals, with the superpotential
$W=G_d(B)$. This can also be regarded as the
theory with gauge group $SU(k)$ which is further gauged
by the $\Z_d$ symmetry generated by
\beq
\gamma:\Phi^a_i\to \e^{2\pi i/kd}\Phi^a_i.
\eeq
It flows in the infra-red limit
to a $\Z_d$ orbifold of the SCFT ${\mathcal
C}_{k,N}(G_d)$ that we introduced in
Section~\ref{subsec:criterion}.
The orbifold is unique since the cohomology ${\rm H}^2(\Z_d,U(1))$
vanishes
\cite{discrete}.

For $d<N$ the FI parameter runs from $r\gg 0$ to $r\ll 0$; for
$d=N$ the FI-theta parameter $t=r-i\theta$ is an exactly marginal
parameter of the IR fixed points; and for $d>N$ it runs from $r\ll
0$ to $r\gg 0$. The effective superpotential $\widetilde{W}$ on
the Coulomb branch can be computed and the vacua can be found,
exactly as before. $\widetilde{W}$ for a fixed trace
$\sum_{a=1}^k\Sigma_a=:\Sigma$ has $n(k,N)$ critical points
(\ref{sigsol}). For each of them, the vacuum equation for the
trace is \beq \Sigma^N=\e^{-t}Z^N(-d\Sigma)^d, \eeq which has
$|N-d|$ solutions at non-zero values and a degenerate solution at
zero. These $|N-d|\times n(k,N)$ non-zero solutions correspond to
massive vacua since $\widetilde{W}$ is non-degenerate there. The
degenerate solution at $\Sigma=0$ corresponds to a superconformal
field theory. These aspects are precisely as in the $k=1$ Abelian
theories \cite{QFTformath}.

To summarize, we learned from
the $U(k)$ linear sigma model that:
\begin{itemize}

\item $d<N$: The non-linear sigma model on the Fano hypersurface
$G_d=0$ in the Grassmannian $G(k,N)$ flows in the infra-red to the
$\Z_d$ orbifold of ${\mathcal C}_{k,N}(G_d)$. It has also
$(N-d)\,n(k,N)$ massive vacua. The IR fixed point is non-singular
if and only if there are no $k$ distinct $N$-th roots of unity
that sum to zero.

\item $d=N$: The K\"ahler moduli space of the Calabi-Yau
hypersurface $G_d=0$ in $G(k,N)$ has one large volume limit,
$n(k,N)$ singular points, and one point that corresponds to the
$\Z_d$ orbifold of ${\mathcal C}_{k,N}(G_N)$. This last point is
also singular if there are $k$ distinct $N$-th roots of unity that
sum to zero. Otherwise, it is a regular theory and the point is
analogous to the Gepner point for the Fermat quintic in $\CP^4$.

\item $d>N$: The $\Z_d$ orbifold of ${\mathcal C}_{k,N}(G_d)$ has
a deformation which drives the theory to non-linear sigma model on
the hypersurface $G_d=0$ of general type in $G(k,N)$. The deformed
theory also has $(d-N)\,n(k,N)$ massive vacua. The UV theory is
non-singular if and only if there are no $k$ distinct $N$-th roots
of unity that sum  to zero.

\end{itemize}
The renormalization group flows and marginal deformations
described  above are consistent with the central charge. The one for
${\mathcal C}_{k,N}(G_d)/\Z_d$ is
\beq \whc=Nk-k^2+1-2N/d=\dim
X_{k,N}(G_d)+2(1-N/d), \eeq
where $X_{k,N}(G_d)$ is the hypersurface $G_d=0$ in the
Grassmannian $G(k,N)$. For $d<N$ and $d>N$, the central charge of
the IR theory is smaller than that of the UV theory. For $d=N$,
the central charge of the CFT ${\mathcal C}_{k,N}(G_d)/\Z_d$ is
the same as the dimension of the Calabi-Yau $X_{k,N}(G_d)$.

We also learned the Witten index of the orbifold theory
${\mathcal C}_{k,N}(G_d)/\Z_d$: It is equal to
the Witten index of the sigma model on $X_{k,N}(G_d)$
minus ({\it resp.} plus) the number of Coulomb branch vacua
$|N-d|n(k,N)$
for $d\leq N$ ({\it resp.} $d>N$).
Note that $n(k,N)$ is given by the formula (\ref{nkN}) in which $\#$
can be set equal to zero
if there are no $k$-distinct $N$-th roots of unity
that sum up to zero.
Thus, the index of the conformal field theory
${\mathcal C}_{k,N}(G_d)/\Z_d$
is, if it is non-singular,
\beq
\Tr\ (-1)^F
=\chi(X_{k,N}(G_d))-{(N-d)(N-1)!\over k!(N-k)!}.
\eeq
The Euler number 
$\chi(X_{k,N}(G_d))$ of the hypersurface can be computed by Schubert calculus
on the Grassmannian.

\subsection*{Duality}

We derive a duality of two dimensional SCFTs, analogous to Seiberg duality in
four-dimensions \cite{seibergduality}. This is
based on the equivalence of the two K\"ahler
manifolds
\be G(k,N)\cong G(N-k,N).
\label{gdual}
\ee
We consider two linear sigma models corresponding to the degree
$d$ hypersurfaces in these (equivalent) spaces: The first is the
$U(k)$ gauge theory with $N$ fundamentals $\Phi_1,...,\Phi_N$ and
one field $P$ in the $\det^{-d}$ representation. These fields are
coupled via a superpotential $W=PG_d(B)$. The second has gauge
group $U(N-k)$, $N$ fundamental fields $\Phi'^1,...,\Phi'^N$ and
$\det^{-d}$ field $P'$. The superpotential is now
$W=P'G'_d(B')$ where $G'_d(B^\prime)$ is the same polynomial as
$G_d(B)$, with the replacement
\be B_{i_1\ldots i_k}=\epsilon_{i_1\ldots i_N}(B')^{i_{k+1}\ldots
i_{N}} \ee
These two theories are equivalent at $r\gg 0$ since they reduce to
the same sigma model.
In the opposite limit $r\rightarrow
-\infty$, the $U(k)$ ({\it resp}. $U(N-k)$)
theory flow to the $\Z_d$ orbifold of
the conformal field theory ${\cal C}_{k,N}(G_d)$
({\it resp}. ${\cal C}_{N-k,N}(G'_d)$).
If $d=N$, the FI-theta parameter is an exactly marginal parameter of
the IR fixed points of both theories. Therefore the equivalence at
$r\gg 0$ means the equivalence of the theories at $r\to -\infty$.
If instead $d<N$, the FI parameters in both theories flow from
$r\gg 0$ to $r\ll 0$. Both theories have $(N-d)n(k,N)$ massive vacua on
the Coulomb branch and one superconformal field theory. Thus, the
equivalence at $r\gg 0$ yields the equivalence of the SCFTs.
For both $d<N$ as well as $d=N$, we found the duality between
the $\Z_d$ orbifold of ${\mathcal C}_{k,N}(G_d)$ and the $\Z_d$ orbifold
of ${\mathcal C}_{N-k,N}(G'_d)$. Unfolding the $\Z_d$ by the quantum symmetry
\cite{unorbifold}, we obtain a duality between the conformal field
theories
\beq
{\cal C}_{k,N}(G_d)\stackrel{{\rm dual}}{\longleftrightarrow}
{\cal C}_{N-k,N}(G'_d).
\label{duality}
\eeq
Equation \eqn{ckNd} confirms that these theories have the same central
charge.
Also, from the discussion in
Section~\ref{subsec:criterion}, we see if one of these SCFTs is singular,
then the other is too with the same
 number of one-dimensional Coulomb branches.
If $d>N$, the relation of the
$U(k)$ and $U(N-k)$ linear sigma models is not strong enough
to prove the duality
but certainly is consistent
with it.
To conclude,
we proved the duality (\ref{duality})
for $d\leq N$ as long as $\whc>0$, 
and we conjecture it for $d>N$.

For the case $N=k+1$, the dual conformal field
theory ${\cal C}_{1,N}(G_d)$ is based on the trivial $SU(1)$ gauge
group and is simply the Landau-Ginzburg model with superpotential
$W=G_d(X)$. This provides the promised, alternative derivation of
the duality of Section~\ref{subsec:confinement} in the case $3\leq
d \leq N$.

It is possible that the conformal field theories ${\mathcal
C}_{k,N}(G_d)$ are mostly new, but it is also possible that some
of them are already known. The latter is the case for the $k=1$
(and hence $N=k+1$) theories: ${\mathcal C}_{1,N}(G_d)$ with
Fermat $G_d$ is the tensor product of minimal models. Possible
candidates for the $k>1$ theories may be found in Kazama-Suzuki
models \cite{KS} --- a class of conformal field theories that can
be realized as gauged Wess-Zumino-Witten models and include the
minimal models as the $SU(2)_{d-2}$ mod $U(1)$ examples. It would
be an interesting problem to find whether there is indeed a
correspondence with such known models, using our knowledge of
${\mathcal C}_{k,N}(G_d)$ such as the central charge, Witten
index, chiral ring, existence of discrete symmetries, etc. In
theories with mass gaps, it was shown in \cite{verlinde} that the
relation (\ref{gdual}) corresponds to the level-rank duality of a
related Wess-Zumino-Witten models. It would be interesting to see
whether the duality (\ref{duality}) of our ``new'' CFTs can be
understood from an alternative point of view.

\subsection{A Test of the IR singularity}

We have seen that the classical mixed Coulomb-Higgs branches that
exist in $U(k)$ linear models for $r<0$ are lifted for all finite
$r$. Nonetheless, in the strict $r\rightarrow \infty$ limit, a
one-dimensional Coulomb branch for the  unbroken $SU(k)$ does
indeed open up provided one can find $k$ distinct $N^{\rm th}$
roots of unity that sum to zero. Examining our list of Calabi-Yau
3-folds, the simplest example where the IR theory is expected to
exhibit a singularity is $X_4\subset G(2,4)$.  This, and related
theories, are the subject of this section. We will make an
independent test of the singularity by examining a dual
description which has nothing to do with the $SU(2)$ Coulomb
branch.

The Grassmannian $G(2,4)$ has the amusing property that it can be
realized as a quadric hypersurface in $\CP^5$. This means that the
$U(2)$ theory with $N=4$ fundamental chirals $\Phi_i$ and a single
$\det^{-4}$ chiral $P$ has a dual Abelian description:
a $U(1)$ gauge theory with six charge $1$ fields which we arrange as
a four-by-four antisymmetric matrix $X_{ij}=-X_{ji}$,
$i,j=1,2,3,4$, and two fields $P_1$ and $P_2$ of charge $-2$ and
$-4$ respectively. A superpotential ${\cal W}=PG_4(B)$ for the
$U(2)$ model is reproduced by a superpotential for the $U(1)$
theory,
\beq W=P_1G_2(X)+P_2G_4(X), \eeq
where
\beq G_2(X)=X_{12}X_{34}-X_{13}X_{24}+X_{14}X_{23},
\label{plucker} \eeq
and $G_4(X)$ is the same as $G_4(B)$ where $X_{ij}$ is inserted in
place of $B_{ij}$. At $r\gg 0$ the $X$ fields span the projective
space $\CP^5$. The superpotential $G_2(X)=0$ cuts out the
Grassmannian $G(2,4)$ in $\CP^5$ (this is the Pl\"ucker relation),
while $G_4(X)=0$ further selects the Calabi-Yau in $G(2,4)$. Thus
the low energy theory agrees with that of the $U(2)$ model at
$r\gg 0$. By analytic continuation the duality must hold for all
values of $t=r-i\theta$.

As discussed in the previous section, in the strict $t\rightarrow
-\infty$ limit, the $U(2)$ model becomes the $\Z_4$ orbifold
of the singular conformal field theory ${\mathcal C}_{2,4}(G_4)$
with central charge $\hat{c}=3$. How can we see this singularity
in the dual $U(1)$ description? At $r\ll 0$, the low-energy
physics of the $U(1)$ theory is governed by a hybrid
Landau-Ginzburg/sigma model on a vector bundle over the weighted
projective space ${\mathbb W}{\mathbb P}^{1,2}$ (a tear drop),
with the base and the fiber spanned by $(P_1,P_2)$ and $X_{ij}$
respectively. The superpotential ensures that the fibre $X$-fields
are massive except over the singular point $P_1=0$ of the base. As
$r\to -\infty$, the Higgs mass blows up and the $U(1)$ vector
multiplet decouples together with one linear combination of
$(P_1,P_2)$. But, most importantly, the size of the tear drop
blows up. This non-compactness in the $U(1)$ model is the sign of
a the singularity in the ${\mathcal C}_{2,4}(G_4)$ conformal
theory.

One can take a closer look at the $U(1)$ theory. It seems
appropriate to focus on the neighborhood of the point $P_1=0$ of
the tear drop, where $P_2$ decouples and $P_1$ remains. Then the
theory is the $\Z_4$ Landau-Ginzburg orbifold with six fields
$X_{ij}=-X_{ji}$ of charge $1$ and one field $P_1$ of charge $2$
with superpotential
\beq W=G_4(X)+P_1G_2(X). \eeq
The R-charges are $1/2$ for $X$'s and $1$ for $P_1$. The model has
central charge $\whc=\sum_{i}(1-q_i)=6\times (1-{1\over
2})+1\times (1-1)=3$. We claim that the IR limit of this LG-model
is dual to the $\Z_4$ orbifold of ${\mathcal C}_{2,4}(G_4)$. The
duality must also hold when the $\Z_4$ orbifold is undone in both
sides. Note that $X_{ij}$ are elementary fields in this dual
theory, while the corresponding operators $B_{ij}$ are composites
in the $SU(2)$ theory. Further, the dual model exhibits no strong
gauge interactions. These features are somewhat reminiscent of the
 $SU(N_c)$ SQCD with $N_f=N_c+1$ flavors in four-dimensions \cite{Seiberg}.
This LG model is obviously singular because the field $P_1$
becomes massless at $X=0$. This is the Higgs branch manifestation
of the singularity of the strong  $SU(2)$ dynamics.

We may also consider the conformal field theory ${\mathcal
C}_{2,4}(G_d)$ for other values of $d$. For $1<d\leq 4$, employing
the $U(2)$ and $U(1)$ linear sigma models as above, we find that
this is dual to the IR fixed point of the Landau-Ginzburg model of
seven variables, $X_{ij}=-X_{ji}$ and $P_1$, with superpotential
\beq W=G_d(X)+P_1G_2(X), \label{gdd} \eeq
where $G_2(X)$ is the quadratic polynomial (\ref{plucker}). Both
theories have central charge $\whc=5-8/d$. (For $d>4$, the linear
sigma models are not strong enough to prove the duality to
(\ref{gdd}) but are certainly consistent with it.) We again find
that the dual theory is singular at $X=0$.

\section{The Glop Transition}
\label{sec:Rodland}

In the previous section, we have shown that the finite $r\ll 0$
phase of the $U(k)$ linear sigma-models is  non-singular. But we
have yet to understand the nature of the low-energy physics at
$r\ll 0$ (with the exception of the models dual to Abelian
theories such as $X_4\subset G(2,4)$). In general \cite{phases},
this regime is described by a gauged Landau-Ginzburg model fibered
over a weighted projective space ${\mathbb W}{\mathbb
P}^{\,q_1,\ldots,q_S}$. These models are typically rather tricky
to study directly. However, one may hope that we could bring the
knowledge learnt in this paper to bear on the problem. It turns
out that we have indeed learnt enough to study the $r \ll 0$ phase
of one further theory associated to the Calabi-Yau 3-folds listed
in Table 2. This is the theory associated to
$X_{1,\ldots,1}\subset G(2,7)$ and it is the subject of this
section.

\subsection{R{\o}dland's Conjecture}

In \cite{rodland} R{\o}dland made a conjecture regarding the
one-dimensional K\"ahler moduli space of a particular (2,2)
superconformal field theory in two-dimensions. He claimed the
moduli space has {\it two} large volume limits for Calabi-Yau
target spaces $X$ and $Y$, defined as follows:

\begin{itemize}

\item $X=X_{1,\ldots,1}\subset G(2,7)$. This is a complete
intersection Calabi-Yau in a Grassmannian $G(2,7)$ defined by 7
generic linear equations of the Pl\"ucker coordinates. We have
already met this object in Section~\ref{subsec:mirror} where we computed the
location of the three singularities in the interior of
the moduli space.

\item $Y=(\mbox{Pfaffian Variety in ${\CP}^{20}$})\cap {\CP}^6$.
The Pfaffian variety ${\rm Pf}(\wedge^2\C^7)$ in $\CP^{20}\cong
\PP(\wedge^2\C^7)$ is defined as the locus of lines $A\in
\wedge^2\C^7$ such that $A\wedge A\wedge A=0$ in $\wedge^6\C^7$.
This means that if we view $A$ as a $7\times 7$ antisymmetric
matrix, then $A$ lies within the Pfaffian variety if each $6\times
6$ sub-matrix has zero Pfaffian (i.e. zero determinant). In other
words, $A\in {\rm Pf}(\wedge^2\C^7)$ if ${\rm rank}(A)$ is
less than the maximal $6$ which, since $A$ is antisymmetric, means
${\rm rank}(A)\leq 4$. Finally $Y$ is defined as the intersection
of ${\rm Pf}(\wedge^2\C^7)$ with a generic 6-plane $\CP^6$ in
$\CP^{20}$.

\end{itemize}

Both $X$ and $Y$ have the same Hodge numbers $h^{2,1}=50$ and
$h^{1,1}=1$ and, in particular, both have a one-dimensional
K\"ahler moduli space. R{\o}dland conjectures that they lie on the
same moduli space \cite{rodland}. As evidence, he studied the
Picard-Fuchs equation for the proposed mirrors of $X$ and $Y$ and
found that the equations are the same. Importantly, $X$ and $Y$
are {\it not} birationally equivalent. This distinguishes this
system from the familiar flop transition. Because of this
important difference, we term the topology changing transition
between $X$ and $Y$ the Grassmannian flop, or glop transition. We
shall now study the glop transition from the perspective of the
linear sigma model.

\subsection{The Genericity Condition}

The linear sigma model for $X=X_{1,\ldots,1}\subset G(2,7)$ is a
$U(2)$ gauge theory with $7$ chiral multiplets $\Phi_i$
transforming in the fundamental representation, and a further $7$
chiral multiplets $P^i$ transforming in the ${\rm det}^{-1}$
representation. The restriction to the compact Calabi-Yau $X$ is
achieved through the introduction of a superpotential
\be W={\mu\over 2}\, A^{jk}_i\,P^i\epsilon_{ab}\Phi^a_j\Phi^b_k
\label{rodsupp} \ee
where $\mu$ is a mass scale and $A^{jk}_i$ are coefficients that
are anti-symmetric in the upper indices: $A^{jk}_i=-A^{kj}_i$. The
F-term equations read
\beq A^{ij}_k\epsilon_{ab}\,\phi^a_i\phi^b_j=0,\ \ \  \ \ \
\ \ p^k A_k^{ij}\phi^b_j=0.\label{F} \eeq
For $r\gg 0$, the D-term equation $D^a_{\,\,b}=0$ (see
\eqn{D}) requires that $(\phi^a_i)$ has rank two.
If the second F-term condition \eqn{F} is strong enough to require
$p^i=0$ then, at energy scales below $\mu$, the theory will flow
to the non-linear sigma-model with target space $X$, defined by
\be X=\Bigl\{\,[\phi^a_i]\in
G(2,7)\,\Bigl|\,A^{ij}_k\phi^1_i\phi^2_j=0,\,k=1,...,7\, \Bigr\}.
\ee
However, we must ensure that the F-term condition \eqn{F} indeed
requires $p^i=0$ for all $[\phi^a_i]\in G(2,7)$. This condition
holds when the two $7\times 7$ matrices $A^i_{\
k}(\phi^a):=A^{ij}_k\,\phi^a_j$ with  $a=1,2$ have a rank $7$
linear combination. This genericity condition on the coefficients
$A_i^{jk}$ is equivalent to the requirement that $X$ is smooth
since, for any $[\phi^a_i]\in G(2,7)$ obeying \eqn{F}, we can find
$7$ normal directions $\delta_lX$  such that
$\delta_l(A^{ij}_k\phi^1_i\phi^2_j)
=A^{ij}_k((\delta_l\phi^1_i)\,\phi^2_j+\phi^1_i\,\delta_l\phi^2_j)$,
regarded as a $7\times 7$ matrix (for indices $l,k$), is of rank
$7$. From now on we pick coefficients $A^{jk}_i$ satisfying this
genericity condition.

\subsection{The Physics at $r\ll 0$}

Let us now turn to the other regime $r\ll 0$. In
this case the D-term condition $D^a_{\,\,b}=0$ requires some $p^i\neq 0$,
so that $p^i$ provide homogeneous coordinates on $\CP^6$. We
define the $7\times 7$ anti-symmetric matrix
\beq A(p)^{ij}:=p^kA^{ij}_k. \eeq
Since $p^i\ne 0$ breaks the $U(2)$ gauge symmetry to $SU(2)$, for
each point $[p^i]\in\CP^6$ the low energy theory consists of an
$SU(2)$ gauge theory with $7$ fundamental chiral multiplets
$\phi^a_i$, with a complex mass term which varies over the
$\CP^6$,
\beq W=\mu \,A(p)^{ij}\phi^1_i\phi^2_j \eeq
To determine the infra-red physics, we can work in a
Born-Oppenheimer approximation, first examining the gauge theory
degrees of freedom and subsequently treating the low-energy
$\CP^6$ sigma-model fields. The character of the gauge theory over
each point $[p^i]\in\CP^6$ depends on the rank of the mass matrix
$A(p)^{ij}$, \beq R:={\rm rank}\, A(p). \eeq We have an $SU(2)$
gauge theory with $R$ massive chiral multiplets and $(7-R)$
massless chiral multiplets. An arbitrary anti-symmetric $7\times
7$ matrix can have rank $0$, $2$, $4$ and $6$. However, the
genericity conditions on $A_i^{jk}$ described above mean that only
$R=4$ and $R=6$ are possibilities. To see this, pick an
orthonormal basis $\{\phi^1,...,\phi^7\}$ of $\C^7$ satisfying
$\sum_{i=1}^7\phi^\mu_i\phi^{\dagger\,\nu}_i=r\delta^{\mu\nu}$,
where $\mu,\nu=1,\ldots 7$. We can split this into three
mutually orthogonal pairs $\{\phi^1,\phi^2\}$,
$\{\phi^3,\phi^4\}$, $\{\phi^5,\phi^6\}$, together with one
remaining orthogonal element $\phi^7$. Each pair defines a point
in the Grassmannian $G(2,7)$. The genericity condition requires
that for each such pair $\{\phi^{2m-1},\phi^{2m}\}$ there exists a
rank $7$ linear combination of $A^i_{\ k}(\phi^{2m-1})$ and
$A^i_{\ k}(\phi^{2m})$. But, in turn, this ensures that the same
linear combination of $\phi^{2m-1}_i$ and $\phi^{2m}_i$ is
non-vanishing when acted upon by  $A(p)^{jk}$ for any $[p]\in
\CP^6$. By construction, we have three such mutually orthogonal
vectors. The remaining vector $\phi^7$ allows us to construct one
more independent pair that is not annihilated by $A(p)$. We
conclude that the genericity condition requires that the rank of
$A(p)$ is at least $4$ at any point $[p]$ of $\CP^6$. This leaves
us with $R=4$ and $R=6$. Let us first look at these in turn,
starting with the $R=6$ domain.

We work in the Born-Oppenheimer approximation, in which the $p^i$
fields are first taken to be fixed. At a point $[p]\in \CP^6$ with
${\rm rank}\, A(p)=6$, the theory of fast variables in the
Born-Oppenhermer approximation is the $SU(2)$ gauge theory with a
single massless chiral multiplet and $6$ chiral multiplets with
complex masses of order $\mu |r|$. From the discussion of
Section~\ref{subsec:cplxmass}, we find that this theory has $3$
vacuum states supported in a halo of radius $\sim \mu |r|$ on the
$SU(2)$ Coulomb branch. In the limit $\mu\rightarrow \infty$, in
which the F-terms \eqn{F} restrict us to the manifold $X$ in the
regime $r\gg 0$, these vacua disappear from sight and decouple
from the theory. We lose the supersymmetric ground states if $[p]$
is deep in the $R=6$ domain. What happens as $A(p)$ degenerates to
rank 4? On the rank $R=4$ locus, we have an $SU(2)$ gauge theory
with 3 massless chiral multiplets and 4 massive chiral multiplets.
Again, from Section~\ref{subsec:cplxmass}, we know that there are
3 ground states, two of which are supported in a halo of radius
$\sim \mu |r|$, while the remaining ground state is localized at
the origin of the Coulomb branch. Hence, as $[p]$ approaches the
$R=4$ locus, one of the three ground states descends from the $\mu
|r|$ halo, to lie in the center of the $SU(2)$ Coulomb branch. We
thus conclude that the low energy theory is localized on the ${\rm
rank}\, A(p)=4$ locus in $\CP^6$.

We may now consider a second stage Born-Oppenheimer approximation,
where we fix a slowly varying profile $p_*$ within the rank four
locus and quantize everything else. Here, ``everything else''
consists of the $SU(2)$ gauge multiplet, the seven fundamentals
$\Phi^a_i$ and the modes of $P^i$ in $\CP^6$ that are transverse
to the $R=4$ locus. Since the $R=4$  locus has dimension three,
there are $6-3=3$ transverse modes, which we denote by $\delta_j
P$, $j=1,2,3$. We may discard the four fundamental chirals that
are massive due to the complex mass matrix $A(p_*)^{ij}$. We
relabel the flavor index so that the first three fundamentals
$\Phi_1,\Phi_2,\Phi_3$ are massless. The chiral multiplets are
thus coupled through the superpotential
\beq W_{\rm fast}=\mu \sum_{i,j=1}^3A(\delta
P)^{ij}\Phi^1_i\Phi^2_j. \label{Wlow} \eeq
Because the rank four locus is defined by
$\delta_1p=\delta_2p=\delta_3p=0$, we may assume the form
\beq A(\delta P)=\left(\begin{array}{ccc}
0&\delta_3 P&\delta_2 P\\
-\delta_3 P&0&\delta_1P\\
-\delta_2P&-\delta_1P&0
\end{array}\right).
\eeq
This theory of fast variables is precisely the theory
\eqn{massiveth} which we considered in
Section~\ref{subsec:confinement}. As we concluded there, it has a
unique supersymmetric ground state with a mass gap. This mass gap
validates the second stage Born-Oppenheimer approximation.

Thus the true low energy degrees of freedom correspond to the motion of
$p_*$ only. We conclude that the low energy theory at $r\ll 0$ is the
non-linear sigma model whose target space is the ${\rm rank}\,
A(p)=4$ locus in $\CP^6$, that is, the Pfaffian variety $Y$. The
Pfaffian $Y$ is a Calabi-Yau threefold and hence the resulting
conformal field theory has $\whc=3$ as expected. In this manner,
the strongly coupled non-Abelian gauge theory has two phases in
which the low-energy physics is described by a sigma-model on
inequivalent Calabi-Yau manifolds, $X$ and $Y$. This confirms the
conjecture of R{\o}dland \cite{rodland}. It is also notable that
we have constructed the linear sigma model for a Calabi-Yau
manifold which is {\it not} a complete intersection of
hypersurfaces in a toric variety. To our knowledge, this is the
first such construction.

\subsection{A Comment on D-brane Categories}

One consequence of the fact that $X$ and $Y$ sit in a common
K\"ahler moduli space is that the categories of topological
B-branes for sigma models on $X$ and $Y$ must be equivalent. In
mathematical terms, $X$ and $Y$ must be derived equivalent, that
is, the associated derived categories of coherent sheaves must be
equivalent, $D^b({\rm Coh}(X))\cong D^b({\rm
Coh}(Y))$.\footnote{This consequence was first brought to our
attention by Edward Witten (April, 2005).} This is particularly
striking since $X$ and $Y$ are not birationally equivalent.
A mathematical proof of the derived equivalence has been given by 
A. Kuznetsov in \cite{AKproof} as an application
of his Homological Projective Duality \cite{Kuznetsov}. The proof
is also given independently in \cite{BC}. It would be very
interesting to construct equivalences of categories from the
physics point of view, as is done for Abelian theories \cite{HHP}.

\subsection{Generalization}

There exist straightforward generalizations of the above story.
For example, we may replace $7$ by $N$. Namely, consider a $U(2)$
linear sigma model with $N$ fundamentals $\Phi^a_i$ and $N$ fields
$P^j$ in the inverse determinant representation, $i,j=1,...,N$
with the superpotential (\ref{rodsupp}) in which $i,j,k$ run over
$1,...,N$. We require the same genericity condition on the
coefficients $A^{jk}_i$: if $\phi^1$ and $\phi^2$ are linearly
independent vectors in $\C^N$, the two $N\times N$ matrices
$A^i_{\ k}(\phi^a):=A^{ij}_k\,\phi^a_j$ with  $a=1,2$ have a rank
$N$ linear combination. Under this condition, the theory at $r\gg
0$ reduces to the non-linear sigma model on a smooth Calabi-Yau
manifold $X_{[N]}$ in $G(2,N)$ defined by $N$ linear equations for
the Pl\"ucker coordinates. The dimension of $X_{[N]}$ is
$d=2N-4-N=N-4$, which is the central charge of the infra-red SCFT,
\beq \whc=N-4. \eeq
At $r\ll 0$, the $p$'s acquire non-zero values, spanning 
$\CP^{N-1}$ and breaking $U(2)$ to $SU(2)$. To study the low energy
physics, we may again work in the Born-Oppenheimer approximation,
where the $SU(2)$ dynamics is classified by the rank $R$ of the
mass matrix $A(p)$. As in the $N=7$ case, the genericity condition
imposes a bound on the rank: For odd $N$ we have $R\geq (N+1)/2$,
while for even $N$ we have $R\geq N/2$. We list below the allowed
ranks for low values of $N$: \beq
\begin{array}{ll}
N=5&R=4\\
N=7&R=6,4\\
N=9&R=8,6\\
N=11&R=10,8,6\\
N=13&R=12,10,8\\
\cdots&\cdots
\end{array}
\qquad
\begin{array}{ll}
N=6&R=6,4\\
N=8&R=8,6,4\\
N=10&R=10,8,6\\
N=12&R=12,10,8,6\\
N=14&R=14,12,10,8\\
\cdots&\cdots
\end{array}
\nn
\eeq
Let us discuss the odd $N$ and even $N$ cases separately.

\noindent
\underline{Odd $N$}\\
[0.2cm] For this case, the Born-Oppenheimer approximation tells us
that the theory localizes on the locus $R\leq N-3$ within 
$\CP^{N-1}$, provided this locus is non-empty. This defines the
Pfaffian variety $Y_{[N]}$ which has dimension $(N-1)-3=N-4$.

The case of $N=9$ is entirely equivalent to $N=7$, since for both
these examples the only allowed non-maximal rank is $R=N-3$. The
second stage Born-Oppenheimer approximation for the $N=9$ theory
again shows that all degrees of freedom, other than the motion in
the $R=6$ locus, are massive. We conclude that the low energy
theory at $r\ll 0$ is the sigma model on the Pfaffian $Y_{[9]}$.
Thus, we again have the glop transition: There is a complex one
dimensional K\"ahler moduli space with two large volume limits;
one corresponds to a Calabi-Yau fivefold in the Grassmannian
$G(2,9)$ and the other to a Pfaffian Calabi-Yau ${\rm
Pf}(\wedge^2\C^9)\cap \CP^8$. There are four singular points in
between.

For $N\geq 11$, there are sub-loci with lower ranks, $R\leq N-5$,
which correspond to the singularities in $Y_{[N]}$. Along such
loci, more than three $SU(2)$ fundamentals become massless and we
expect additional low energy degrees of freedom. The low energy
theory is not a simple non-linear sigma model, which is anyway
ill-defined, but a theory with extra degrees of freedom along the
$R\leq N-5$ loci. These extra degrees of freedom must be
responsible for the ``quantum resolution'' of the singularity of
$Y_{[N]}$. \footnote{Alexander Kuznetsov pointed out that the
Pfaffian Calabi-Yau $Y_{[N]}$ are singular, with the exception of
$N=7,9$, and suggested that one should consider certain
non-commutative resolution of this singular variety when the
derived category is discussed \cite{AKproof}.}

The case $N=5$ is special in that there is no point with
$R=N-3$.
Within the Born-Oppenheimer approximation,
we only find two vacuum states supported in a region
away from the origin
in the $SU(2)$ Coulomb branch.
In particular, we cannot identify the low energy theory that may flow to
a $\whc=5-4=1$ SCFT which we do expect for any value of $t=r-i\theta$.
This means that the Born-Oppenheimer approximation
is not valid in this case.

\noindent
\underline{Even $N$}\\
[0.2cm]
We first note, following the discussion of Section~\ref{subsec:cplxmass},
  that the ``decoupling'' of flavors with large complex masses
is qualitatively different from the odd $N$ case. The states that
decouple are not supported away from the origin, but rather spread
over a wide region in the $SU(2)$ Coulomb branch. This leads to
 the singularity which we expect in the strict $t=-\infty$
limit. Thus, we do not expect the clean localization of low
energy degrees of freedom at the loci where $A(p)$ has small rank.

Suppose we try to proceed by ignoring this subtlety. Then we would
have to conclude that the low energy theory localizes on the
$R\leq N-4$ locus which we denote by $Z_{[N]}$. This locus has
dimension $N-7$. The central charge for the degrees of freedom
within $Z_{[N]}$ is bounded to be $\leq N-7$, which falls  short
of the expected central charge $\whc=N-4$ by at least three. This
means that we need extra massless degrees of freedom from the
modes transverse to $Z_{[N]}$. In other words, there is no
separation of energy scales and the Born-Oppenheimer approximation
is out of question.

To conclude, the Born-Oppenheimer approximation is not valid for
the case of even $N$, and we do not yet know the nature of the
$r\ll 0$ phase.

\section*{Acknowledgement}
We would like to thank Victor Batyrev, Andrei Caldararu,
Charles Doran, Nick Dorey, Ionut Ciocan-Fontanine,
 Sean Hartnoll, Bumsig Kim, Alexander Kuznetsov,
 David Morrison, Sunil Mukhi, Erich Poppitz,
Cobi Sonnenschein, Duco van Straten and Edward Witten
for several useful discussions and correspondences.

Both authors
are grateful to the Research Institute for Mathematical Sciences
and the Yukawa Institute for Theoretical Physics,
Kyoto University,
for their kind support during the visit in the summer of 2004
where a major part of this work was done, and to the
Institute for Theoretical Physics in Amsterdam for the
2006 Summer Workshop which motivated us to write up a paper.
K.H. also thanks
the Banff International Research Station where he was introduced
to R{\o}dland's work by Duco van Straten,
and the Fields Institute and the Perimeter Institute
for
useful discussions with participants to the 2004-5
 Thematic Program on the Geometry of String Theory.

K.H. is supported also by
NSERC and Alfred P. Sloan Foundation.
D.T. is supported by the Royal Society.

\end{document}